\documentstyle[aps,eqsecnum,epsf,epsfig]{revtex}

\begin{document}
\baselineskip=16pt
\draft
\title{Quantum collisions of finite-size ultrarelativistic nuclei.}
\author{ A. Makhlin\footnote{e-mail: makhlin@nscl.msu.edu}}
\address{Department of Physics and Astronomy and National
Superconducting Cyclotron Laboratory, \\ Michigan State
University, East Lansing, Michigan 48824}
\date{April 3, 2001}
\maketitle

\begin{abstract}
We show that the boost variable, the conjugate to the coordinate
rapidity, which is associated with
the center-of-mass motion, encodes the information about the
finite size of colliding nuclei in a Lorentz-invariant way.
The quasi-elastic forward color-changing scattering between the
quantum boost states rapidly grows with the total energy of the
collision and leads to an active breakdown of the color coherence
at the earliest moments of the collision. The possible physical
implications of this result are discussed.
\end{abstract}


\section{Introduction}

It is commonly accepted that on the scale of the strong interaction,
which is responsible for nuclear integrity and compactness, the large
nuclei have a macroscopically finite size and a well-defined
boundary.\footnote{By the finite size, we mean the size which is
measured by means of the strong interaction of two nuclei. If the
primary interaction were electromagnetic (as is in the {\em ep}-or {\em
eA}-processes), then the whole concept of a finite size would become
doubtful.} This size can be physically measured in the rest frame of a
nucleus, and it undergoes the Lorentz contraction in the moving frame
without any physical limitations (as is required by special
relativity). In this paper, we suggest to take this fact as a
guideline, and explore the consequences of the finite size of the
nuclei for the {\em quantum process} of their collision at
ultrarelativistic energies. Of these consequences, the most important
is the change of the symmetry: The incoming nuclei are prepared in a
homogeneous space having a given energy and momentum.  The fixed space-
time point of the first interaction corrupts the initial symmetry, and
enforces a different choice of the conserved quantum numbers for the
later stages.  Of the ten symmetries of the Poincar\'{e} group, only
rotation around the  collision $z$-axis, Lorentz transformation along
it,  and the translations in the transverse $x$ and $y$-directions
survive. Therefore, it is profitable to choose, in advance, the set of
normal modes which have the symmetry of the localized initial
interaction and carry quantum numbers adequate to this symmetry. These
quantum numbers are the transverse components, ${\vec p}_t$, of
momentum and the boost, $\nu=p^0z-p^zt$, of the particle (which is
associated with the center-of-mass motion and replaces the component
$p^z$ of its momentum).

These geometric considerations can be reinforced  by the quantum
mechanical arguments. Indeed, from the perspective of an external
observer, the first thing that happens during the collision is a
precise measurement, by means of the strong interactions, of the
collision coordinate within a very short time interval. Therefore,
statistically, by the uncertainty principle, the secondaries with
any conceivable momentum $p_z$ can be detected after collision. This is
a well-known scheme of the Heisenberg microscope. The higher resolution
we want to achieve, the larger must be the energy resources of the
microscope. In the textbook example of the electron probed by the
photon, the electron receives energy from the hard photon. In nuclear
collisions, both the kinetic energy of the nuclei, and the energy of
the compression of the Lorentz-contracted nuclei are used for the
purpose of a precise measurement of the coordinate. An internal
observer that penetrates the future of the collision with the nuclei
will see a violently expanding matter around him. The two viewpoints
perfectly complement each other. The short scales of primary
interaction provide a sufficient motivation to use the wedge dynamics
which describes the fields inside the future domain of the ``wedge''
$\tau^2=t^2-z^2>0$, and employs the ``proper time'' $\tau$ as a
Hamiltonian time of the evolution and the coordinate  rapidity $\eta$
as a longitudinal coordinate \cite{gqm,geg}. The infamous rapidity
plateau persistently observed in high-energy nuclear collisions
strongly supports this picture.

The approach advocated in this paper explicitly incorporates the
macroscopic finite size of the interacting objects into the quantum
theory of the earliest stage of the collision. We assume that there is
no measurable gluon fields outside the large stable nuclei.
Consequently, the time moment and the $z$ coordinate, along the
collision axis, of the first interaction are defined with the accuracy
of at least $\sim 0.01fm$, which is both the size of a Lorentz-
contracted individual nucleon, and the characteristic scale of color
correlation in the $z$ direction before the collision. The full size of
the Lorentz-contracted gold nucleus at the energy $\sim 100GeV$ per
nucleon is $\sim 0.1fm$. We show, that despite an almost infinite
Lorentz contraction and the quantum nature of the interaction process,
the information about the finite size of the incoming nuclei does not
fade away.  It remains clearly identifiable in terms of the properly
chosen Lorentz-invariant variable, {\em the boost}, which is
associated with the center-of-mass motion. Thus, it is possible to
describe the collision of the two nuclei staying on the same physical
ground in any reference frame, either in the reference frame of one of
the nuclei, or in the laboratory frame where both nuclei move almost at
the speed of light.

The fact that nuclei have finite size is intimately connected with the
gauge nature of the strong interactions. Therefore, when addressing the
problem of interaction of the two compact nuclei, we must refer to the
properties of the vector gauge fields.\footnote{Addressing the issue of
interaction of finite-size nuclei, one should keep in mind the source
of the major difference between QED and QCD phenomena. The local gauge
symmetry of QED can be extended to a global gauge symmetry which
generates the conserved gauge-invariant global quantum number, the
electric charge, which can be sensed at a distance. The proper field of
an electric charge is the main obstacle for the definition of its size.
On the other hand, the radiation field of QED appears as a result of
the changes in the extended proper fields of accelerated charges, and
one can physically create such an object as a front of  electromagnetic
wave. In QCD, the local gauge invariance of the color group cannot be
extended to is global version that would correspond to a gauge-
invariant conserved charge. Hence, we can readily define the size of
the colorless nucleus, but we cannot create a front of color radiation
in the gauge-invariant vacuum. These two properties of QCD both work
for us. They allow one to use the Lorenz contraction to localize the
primary domain of the collision and thus, to impose the classical
boundary conditions on the color fields at later times. The existence
of the collective propagating quark and gluon modes at the later times
is the conjecture that has to be verified by the study of heavy ion
collisions.} The colored sources of these fields must be located inside
the nuclei and they can be physically resolved only after the two
nuclei overlap. This is the only assumption we make regarding the
internal structure of a nucleus. By all means, location of a material
object inside a nucleus implies that its center-of-mass should move
with this nucleus without crossing its boundary. Therefore, before the
collision, it is natural to characterize such an object by its center-
of-mass, i.e., by its boost $\nu$. The valence quarks are the first
candidates to be considered in this manner. In this sense, we follow
the idea of McLerran-Venugopalan model \cite{McLerran1} in the form
given by Kovchegov and Mueller \cite{Kovchegov}. However, we do not try
to populate the nuclei with the wee partons. We think that they are
{\em gradually created} in the course of collision \cite{QGD,tev}.

The framework of wedge dynamics also offers a unique opportunity to
avoid various technical problems encountered when the moving at the
speed of light nuclei, $V_z=\pm c$, are taken as the first
approximation \cite{McLerran1,QGD}. This state cannot be reached as a
continuous limit of $V\to c$ and a significant effort has been made to
smooth out the singular behavior of quantum fields at $V=c$
\cite{tev,McLerran2,Lam1}. The wedge form of Hamiltonian dynamics is
free of this difficulty. Furthermore, the gauge $A^\tau=0$ of the wedge
dynamics can be fixed completely. Hence, the transverse and
longitudinal fields are well separated and the gluon propagators of
wedge dynamics have no spurious poles that can stimulate a singular
behavior of scattering amplidudes \cite{geg}. In this framework, one
can use the same dynamics and the same gauge for the description of
both incoming nuclei and the products of their reaction \cite{QGD,tev},
thus avoiding all glitches of the ``on-line'' changing the gauge and
redefinition the states \cite{KMW}.

Below, we concentrate on a specific interaction in the expanding system
that emerges in ultrarelativistic nuclear collisions. It is mediated by
the longitudinal part of the gluon field.\footnote{The division of the
gauge field into the longitudinal and transverse parts can be done only
with respect to the property of propagation: transverse fields are
emitted and then propagate being limited in space-time by the light-
cone boundaries, while the longitudinal fields are simultaneous (in
terms of the Hamiltonian time) with their sources. In QCD, this scheme
can be practically implemented only in the framework of perturbation
theory, which is assumed throughout this paper.} It seems to be the
leading one at the earliest moments of  the collision of the two
nuclei, and to result in the intensive color exchanges even in quasi-
elastic subprocesses. Eventually, these exchanges must cause an active
breakdown of the fragile color coherence of the colliding nuclei and
stimulate intense color radiation. The rate $\sigma_1$ of these color
exchanges between the quantum boost states appears to be large at the
earliest moments of the collision, and it grows as $log^2E$ with the
total energy $E$. This major result of this paper is given by
Eq.(\ref{eq:E2.13e}). The $log^2E$ dependence of the rate on the total
energy of the collision resembles the obtained in the early sixties
estimate on the maximal rate at which the total cross section may grow
with the energy. It is known as the Froissart bound, and a steady
growth of the total cross section is indeed observed in the
proton-proton collisions.

Originally, the Froissart bound was derived in the scope of the
axiomatic field theory, a powerful approach based on the most general
requirements, like Lorentz invariance, causality, unitarity,
completeness of the basis of physical states and the cluster
decomposition principle (see Refs.\cite{Weinberg} for the details).
Since the perturbation theory (usually {\em in a given order}) can lead
to an anomalously large total cross section (and thus to apparently
violate unitarity) it is said that the perturbative total cros-section
requires unitarization. Recently, this problem received a vigorous
attention in connection with the evolution equations for large nuclei
at low $x_F$ \cite{Kovner,KMW}. A physical protection from an excessive
growth of cross section due to collinear problems was offered in Ref.
\cite{tev}. From this standpoint, one can infer that the result
(\ref{eq:E2.13e}) of this paper indeed complies with the unitarity.
Though this issue has to be studied in more details, we suggest a
plausible simple physical arguments below.

The axioms of unitarity and completeness clearly are not truly
ndependent. Discussion of any issue related to unitarity requires  that
the spaces of the initial and final states are completely specified.
Physically, this means that the measurement is not accomplished until
its products are analyzed. What the particular states are, depends on
the detectors that resolve these states. In nuclear collisions, one
cannot rely on the conventional ``external'' distant detectors. The
role of the detectors for the earliest subprocesses (which only very
tentatively can be viewed as the independent acts of scattering) is
played by the subsequent interactions.  The next-to-best thing one can
do is to try to answer the following question. Let the  fields excited
at the beginning of  the collision are expanded over a system of states
characterized by some quantum numbers. Let two such states interact.
What is the rate of these interactions? The answer will be related to
the two main problems of ultrarelativistic heavy ion collisions. First,
the known rate of the primary interaction will help to estimate the
entropy production. At this point, the explicit knowledge of the final
states is imperative, because the entropy is the number of the excited
degrees of freedom. Second, it will be directly connected to the total
cross section. Indeed, if the fields change their colors during the
time $\sim 1/E$ with sufficient probability, then the nucleons will
lose their coherence and fall apart. A new composition of hadrons will
be created with the probability one, and it does not really matter how
the interacting states are chosen. This argument has been tested long
ago: the total cross section of the $e^+e^-$-annihilation into hadrons
coincides with the cross section of the process  $e^+e^-\to q\bar{q}$.
One of the recently studied examples is the interaction of the
eikonalized quarks or gluons \cite{zahed}. In this paper, for the same
purpose, we consider the ``natural'' states of the wedge dynamics,
deliberately leaving the key question of {\em what} interacts at the
very beginning of the collision open. We find that, because the states
of wedge dynamics carry internal currents in the coordinate rapidity
direction, there exists a specific contact interaction of these
currents which grows when $\tau\to 0$, and leads to the amplitude of
interaction, proportional to $\log~E$. [The contact term in the gluon
propagator has been singled out in the course of the complete fixing of
the gauge $A^\tau =0$, and its main effect is confined to the nearest
vicinity of the light wedge, $\tau=0$, where the boundary conditions
that fix the gauge are imposed.] If the QCD indeed falls under a
jurisdiction of the axiomatic field theory (which by no means is self-
evident), then our perturbative result, which exactly reaches the
Froissart bound, may point to the major physical mechanism that
triggers the scenario of ultrarelativistic heavy ion collisions.

The paper is organized as follows. In section \ref{sec:SN1} we
introduce the variables of wedge dynamics and clarify the physical
meaning of the boost in classical and quantum contexts. In section
\ref{sec:SN2} we use the boost states to estimate the amplitude of
forward scattering with color transfer at the earliest moments of the
collision, paying attention to the contact interaction in the expanding
system. In Appendix A, we demonstrate that the contact term has no
counterparts, and that  the standard Coulomb-type terms are still there
in the propagator. They are somewhat modified, just in a way which one
could expect on purely physical grounds. In Appendix B we show, that
the contribution of the other terms into the forward scattering
amplitude is subleading.

\section{Classical and quantum particles in wedge dynamics}
\label{sec:SN1}

\renewcommand{\theequation}{2.\arabic{equation}}
\setcounter{equation}{0}

In this section, we address the basic connection between the
classical and quantum aspects of the interaction of compact
relativistic objects, in order to prepare the stage for a more
involved analysis of the interaction picture. First, we discuss
the role of the classical Lorentz boost as a natural variable
which, by its origin, is closely related to the finite size.
Second, we review the meaning of the boost as a quantum number,
and establish its connection with the classical boost. Finally, we
show that the genuinely classical distribution of the boosts in
stable nuclei before the collision plays a role as the initial
data for the primary quantum interactions between nuclei.

\subsection{Introducing the variables. }
\label{sec:Sb1.1}

The wedge form of relativistic dynamics works inside the future
domain of the hyperplane $t=z=0$ (light wedge) were the two
finite-size ultrarelativistic objects touch each other for the
first time. The natural coordinates inside this domain are
parameterized by the proper time $\tau$ and the rapidity
coordinate $\eta$,
\begin{eqnarray}
t=\tau ~\cosh\eta,~~~~z=\tau ~\sinh\eta~.
\label{eq:E1.1}\end{eqnarray}

In terms of these variables, the action for a classical particle is
\begin{eqnarray}
S=\int L d\tau =-m~\int ds =-m~\int\sqrt{1-v^2}~d\tau = -m~\int
d\tau\sqrt{1-\tau^2\dot{\eta}^2-\dot{\vec r}^2},
\label{eq:E1.2}\end{eqnarray}
where $v^2\equiv\tau^2\dot{\eta}^2+{\dot{\vec r}}^2$
is the spatial velocity squared, and the dot means derivative over
the (Hamiltonian)  time $\tau$.\footnote{Following a tradition, we
use the Greek indices for the four-dimensional vectors and tensors
in the curvilinear coordinates ($\eta$ is an exception, it always
stands for the rapidity direction), and the Latin indices from $a$
to $d$ for the vectors in flat Minkowsky coordinates. We use
Latin indices from $r$ to $w$ for the transverse $x$- and
$y$-components ($r,...,w=1,2$), and the arrows over the letters
to denote the two-dimensional vectors, {\em e.g.},
${\vec k}=(k_x,k_y)$, $|{\vec k}|= k_{t}$. The Latin indices
from $i$ to $n$ ($i,...,n=1,2,3$) will be used  for the
three-dimensional internal coordinates $u^i=(x,y,\eta)$ on the
hyper-surface $\tau=const$.} The canonical momenta of this particle,
\begin{eqnarray}
p_\eta\equiv\nu={\partial L\over\partial\dot{\eta}}=
{m\tau^2\dot{\eta}\over\sqrt{1-v^2} },~~~~~
{\vec p}={\partial L\over\partial\dot{\vec r}}=
{m \dot{\vec r}\over\sqrt{1-v^2}}~,
\label{eq:E1.3}\end{eqnarray}
are conserved by virtue of the equations of motion.
The Hamiltonian is of a standard relativistic form,
\begin{eqnarray}
H=\nu\dot{\eta}+{\vec p}\cdot\dot{\vec r}-L=
{m\over\sqrt{1-v^2}}~,
\label{eq:E1.4}\end{eqnarray}
which, after excluding the velocities, can be rewritten
in terms of the canonical momenta,
\begin{eqnarray}
H=\sqrt{m^2 + {\vec p}^2 +{\nu^2\over\tau^2}}.
\label{eq:E1.5}\end{eqnarray}
The useful relations of geometric origin, which will be
often referred to later on, are
\begin{eqnarray}
p^\eta=-{1\over\tau^2}~ p_\eta=-\bigg({\sinh\eta\over\tau}~p^0-
{\cosh\eta\over\tau}~p^3\bigg)=
-{m_t\over\tau}\sinh(\eta-\theta),\nonumber\\ H=\cosh\eta~ p^0-
\sinh\eta~ p^3 =m_t\cosh(\eta-\theta),
\label{eq:E1.6}\end{eqnarray}
where  $m_t^2=m^2+p_t^2$,
$~p^0=m_t\cosh\theta$, and $p^3=m_t\sinh\theta$ are the Cartesian
momenta. Therefore, the boost,
\begin{eqnarray}
\nu = p_\eta= \tau m_t \sinh(\eta-\theta)=x^3p^0-x^0p^3
\equiv p^0(z-V_z t),
\label{eq:E1.7}\end{eqnarray}
is related to the center-of-mass coordinate. According to
Eq.(\ref{eq:E1.5}),
the quantity $\nu/\tau$ plays a role as a local longitudinal momentum.

The Hamilton-Jacobi equation for the classical action of a
particle reads as
\begin{eqnarray}
{\partial S\over\partial\tau}+
\sqrt{{1\over\tau^2}~\bigg({\partial S\over\partial\eta}\bigg)^2
+\bigg({\partial S\over\partial{\vec r}}\bigg)^2+m^2}=0~.
\label{eq:E1.8}
\end{eqnarray}
It allows for the separation of variables and has a solution
\begin{eqnarray}
S=\nu\eta +{\vec p}\cdot{\vec r}-
\int\sqrt{m_t^2+{\nu^2\over\tau^2}} d\tau =\nu\eta +{\vec p}\cdot{\vec
r}-\sqrt{m_t^2\tau^2 +\nu^2}+ \nu {\rm Arsh}{\nu\over m_t\tau}~.
\label{eq:E1.9}\end{eqnarray}
In a quantum context, this action serves as the
phase of a semi-classical wave function, $\psi\sim e^{iS}$, with
the quantum numbers $\nu$ and ${\vec p}$, either when $\nu \gg \tau
m_t$ or when $\tau m_t \geq \nu  $. An isolated solution with the
not separated variables is
\begin{eqnarray}
S={\vec p}\cdot{\vec r}-m_t \tau \cosh(\eta-\theta)~.
\label{eq:E1.10}\end{eqnarray}
It corresponds to a plane wave, and its parameter, the
(momentum) rapidity $\theta$,
is not a canonical momentum.

\subsection{Classical trajectories. The physical meaning of
the boost $\nu$. }
\label{sec:Sb1.2}

In order to understand the physical meaning of the boost variable
$\nu$, the canonical conjugate to the rapidity $\eta$, one has to
figure out how it enters the classical equations of motion.
According to the Jacobi theorem, the action $S(x_n,a_n)$, known as
a function of coordinates $x_n$ and arbitrary constants $a_n$,
allows one to find an additional set of the conserved quantities,
$ ~\partial S/ \partial a_n=b_n $. While the constants
$~a_n=\partial S/ \partial x_n$ usually are the canonical
momenta corresponding to the cyclic coordinates and conserved due
to the equations of motion, as in Eq.~(\ref{eq:E1.9}), the
constants $b_n$ appear to be the initial coordinates. Applying
the Jacobi theorem to the action (\ref{eq:E1.9}), and choosing
the constants in such a way, that at $\tau=0$ we have $x=x_0$, and
that at $\tau\to\infty$ we have $\eta =\theta$, we obtain the
equation of the particle trajectory,
\begin{eqnarray}
x(\tau)-x_0={p_x\over m_t^2}~
(\sqrt{\tau^2 m_t^2 +\nu^2}-|\nu|)~ ,\nonumber\\
\eta(\tau)-\theta= - {\rm Arsh}{\nu\over m_t\tau}~.
\label{eq:E1.11}\end{eqnarray}
Despite their unusual appearance, these two equations parameterize
a straight line, as it should be for the free motion of a point-like
particle. Let us rewrite the second of equations (\ref{eq:E1.11})
in two ways,
\begin{eqnarray}
m_t\tau \sinh[\eta(\tau)-\theta]=\nu=
zp^0 -tp^z ~\to ~ m_t z_\ast~,
\label{eq:E1.12}\end{eqnarray}
and
\begin{eqnarray}
m_t\tau \cosh[\eta(\tau)-\theta]= tp^0-zp^z=
\sqrt{\tau^2 m_t^2 +\nu^2}~\to ~m_t t_\ast~,
\label{eq:E1.13}\end{eqnarray}
where the arrows point to the special choice of the reference frame
with $\theta=0$.\footnote{ We consider the physical  design of the
nucleus as almost static, and neglect the possible velocity  $V^z_\ast$
of the nuclear constituent in the nuclear rest frame. In any case, it
cannot be large without undermining the alleged stability of the
nucleus. The origin of the transverse mass may be different. It
includes both the Lagrangian mass and the ``adjoint mass'' due to the
transverse momentum. Inside a stable nucleus, the momenta most
probably characterize the standing waves which are not likely to be too
short, if the nucleus is in the ground state.} Then the first of the
equations
(\ref{eq:E1.11}) becomes
\begin{eqnarray}
x(\tau)-x_0= {p_x\over m_t}
\bigg(\sqrt{\tau^2 + {\nu^2 \over m_t^2 }}
-{|\nu|\over m_t}\bigg) ~\to ~ {p_x\over m_t}( t_\ast-|z_\ast|)~,
\label{eq:E1.14}\end{eqnarray}
obviously satisfying the required boundary condition at $\tau=0$.  Now,
it is easy to understand that the quantity $\nu / m_t$ is the $\tau$-
independent coordinate $z_\ast$ of the particle in the co-moving frame.
By the definition, this quantity is Lorentz-invariant: the boost $\nu$
is the same in all Lorentz frames. The Cartesian form (\ref{eq:E1.14})
of the trajectory is obviously continued to all quadrants of the $tz$
plane. This classical definition of the boost is fairly operational
but, as the reader may notice, it requires that the base world line
(plane), from which the coordinate $z_\ast$, is measured is explicitly
chosen. For the two colliding nuclei, it is natural that the base lines
(corresponding to the rapidities $\pm Y$) go through the point $t=z=0$,
where the nuclei touch each other by their surfaces. If the nuclei have
radius $R$ and are built from the fragments of the (transverse) mass
$m_t$,  then the  boosts for the right-moving nucleus will be in the
range  $-2m_tR < \nu<0$, and in the range $0< \nu < 2m_tR$ for the
left-moving one. There is no contradiction with quantum mechanics  at
this point, since the nuclei are  {\em macroscopic} stable  objects
which can be kept under non-destructive control (in their co-moving
reference frames) before the collision. Asymptotically, they have the
well defined rapidities $\theta=\pm Y$, which can be also measured
classically, without any contradiction with the anticipated uncertainty
relation, $\Delta\nu\Delta\eta\geq 1$. Indeed, the boosts $\nu\approx
m_t z_\ast$ are measured {\em inside} the nuclei, while the
measurements of the velocities of the nuclei is performed by external
devices. Therefore, the boost variable is indeed perfectly suited for
the description of the finite size objects. If the relative boosts of
all constituents do not change in the course of the interaction, then
the object remains unaltered in its (possibly new) rest frame.

\begin{figure}[htb]
\begin{center}
\mbox{
\epsfig{file=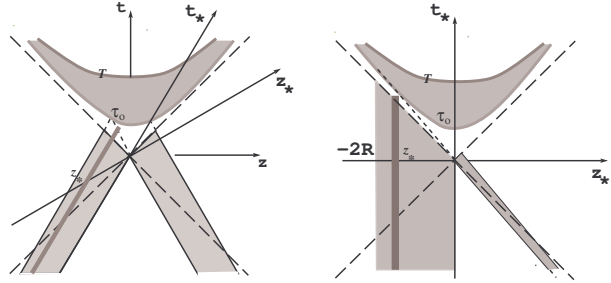,height=1.5in,bb=100 560 530 755}
}
\end{center}
\caption{Geometry of a nucleus-nucleus collision in the center-of-mass
reference frame (left) and in the rest frame of one of the nuclei
(right) The dark gray lines correspond to a semi-classical boost state
in the right-moving nucleus before the collision. }
\label{fig:figb}
\end{figure}

As the matter of fact, the boosts provide an {\em invariant measure} of
the distribution of the constituents of the compact objects. The
picture of rectilinear trajectories holds outside the light wedge also.
Therefore, the classically prepared distribution of the boosts is
resolved as the distribution of the further interacting quantum states
with the given boosts, when two such objects collide. Though
Eq.~(\ref{eq:E1.12})  expresses the boost $\nu$ via the invariant $m_t$
and distance $z_\ast$, in a quantum picture, the boost $\nu$ is an
independent conserved additive quantum number.

For  isolated point-like (and thus, structure-less) objects, the
practical measurement of the boost requires that the rapidity
$\eta(\tau)$ is measured at two time moments along the same
trajectory. Then, solving the system of two equations
(\ref{eq:E1.12}), one finds $\nu$ and $\theta$, the boost and the
asymptotic rapidity of the particle. It is unrealistic to perform
such measurements with sufficient accuracy in the asymptotic
domain of the macroscopically large $\tau$. Unlike the case of
the macroscopic finite size object, this kind of measurement does
meet quantum-mechanical obstacles.

\subsection{The boost $\nu$ in quantum context. }
\label{sec:Sb1.3}

The quantum-mechanical measurement of the boost $\nu$ is very
similar to the measurement of a usual momentum and relies on the
definition of the operator of the boost,
\begin{eqnarray}
\hat{\nu}=-i~{\partial\over\partial\eta}~,
\label{eq:E1.15}\end{eqnarray}
as the operator of translations in the $\eta$-direction. Then it
becomes evident, that a simultaneous  measurement of coordinate
$\eta$ and momentum $\nu$ is limited by the uncertainty relation,
\begin{eqnarray}
\Delta \nu~\Delta\eta ~\geq 1~. \label{eq:E1.16}\end{eqnarray}

In the field-theory formulation, the boost operator is given by the
generator of the Lorentz rotations in the $tz$ plane. In the
internal geometry of wedge dynamics, the boost operator is
given by the $\tau\eta$ component of the energy-momentum tensor.
The boost of the quantum field at the proper time $\tau$,
\begin{eqnarray}
\nu=\int_{\tau=const}T^{\tau\eta}(x) \tau d\eta d^2\vec{r}
=\int d\Sigma_\mu M^{\mu 03}(x)~,
\label{eq:E1.17}\end{eqnarray}
(where $M_{\mu\nu\lambda} =x_\nu T_{\mu\lambda}-x_\lambda T_{\mu\nu}+
S_{\mu\nu\lambda}$ is the usual angular momentum tensor) is the
integral of motion corresponding to the translation symmetry (Lorentz
rotation) in $\eta$ direction. The quantum states with the given
boost $\nu$ are the eigenstates of the operator (\ref{eq:E1.15}),
and their eigenfunctions depend on $\eta$ as $e^{i\nu\eta}$.
The full wave function of a scalar particle with the boost $\nu$
and the transverse momentum ${\vec p}$ is the solution of Klein-Gordon
equation with the separated variables $\tau$, $\eta$ and ${\vec r}_t$,
\begin{eqnarray}
\psi^{(+)}_{{\vec p},\nu}(x)={e^{-\pi\nu/2}\over 2^{5/2}\pi }
H^{(2)}_{-i\nu} (m_{t}\tau)
e^{i\nu\eta +i{\vec p}{\vec r}}~.
\label{eq:E1.18}\end{eqnarray}
It is normalized on the hypersurfaces $\tau=const$,
\begin{eqnarray}
\int_{\tau=const}\psi^{\ast}_{\theta', {\vec p}'}(x)
~i{\stackrel{\leftrightarrow}{\partial \over \partial \tau}}~
\psi_{\theta, {\vec p}}(x)~\tau d\eta d^2{\vec r}=
\delta (\theta -\theta')\delta ( {\vec p}- {\vec p}')~.
\label{eq:E1.18a}\end{eqnarray}
This equation  normalizes the measurements performed by an array of the
detectors moving with all possible velocities. At any particular time
of the Lorentz observer, this array even does not cover the whole
space.

At large $\nu\gg 1$, and $\nu>m_t\tau$, which is relevant to the
earliest stage, the asymptotic of this solution is semi-classical,
\begin{eqnarray}
\psi^{(+)}_{{\vec p},\nu}(x) \approx { e^{i\pi/4}\over 4\pi^2 }
~{ e^{i\nu\eta +i{\vec p}{\vec r}}\over [m^2_{t}\tau^2+\nu^2]^{1/4}}~~
e^{-i\sqrt{m^2_{t}\tau^2+\nu^2} +i\nu{\rm Arsh}(\nu/m_{t}\tau)}
\propto e^{iS}~,
\label{eq:E1.19}\end{eqnarray}
clearly indicating that at the small time $\tau$ the quantum particle
with the finite boost $\nu$ continues to follow its classical
trajectory, since its classical action is large. Indeed, the
surface of the light wedge, everywhere except for its vertex,
corresponds to $\eta\to\infty$.

The wave functions (\ref{eq:E1.18}) are connected, by means of
Fourier transform, with the plane-wave solutions,
\begin{eqnarray}
\omega^{(+)}_{{\vec k},\theta}(x)
= \int_{-\infty}^{+\infty}  {d\nu \over (2\pi)^{1/2}i} e^{-i\nu\theta}
\psi^{(+)}_{{\vec k},\nu}(x)~={1\over 4\pi^{3/2} k_{t}}
e^{-ik_{t}\tau\cosh (\theta-\eta) +i{\vec k}{\vec r}}~.
\label{eq:E1.20} \end{eqnarray}
The saddle point of the Fourier transform (\ref{eq:E1.20})
(or its inverse)  is located at the value of $\nu$ (or $\theta$)
defined by the relation, $~\nu=\tau m_t\sinh(\theta-\eta) ,~$
corresponding to the classical definition (\ref{eq:E1.7}) of
the boost. One can easily see that these wave functions also
are semi-classical with the action (\ref{eq:E1.10}), and have
a usual momentum (or the rapidity $\theta$) as a quantum number.
These states become localized in rapidity $\eta$ at later times,
$\tau m_t\gg 1$, and namely these states are most likely to be
detected by the expanding collective system.

The key element of the suggested approach is that the Lorentz-invariant
boost states, which are independently prepared in the two approaching
nuclei, begin to interact as the quantum states only when the nuclei
overlap. At this moment, the positions of the nuclei constituents
(classical boosts, which describe the elementary constituents of the
nuclei even outside the light wedge) are translated into the quantum
numbers, which define the periodicity of the wave functions in the
coordinate rapidity direction. It is evident, that at the earliest
times, the distortion of the initial geometric picture should be only
minimal. Therefore, it will be a sufficient approximation to study the
transitions into other boost states, and we stay within this
approximation until the end of this paper. The rate at which these
early distortions develop appears to be quite large.

The dynamics of boost states preserves the {\em invariant information}
about the finite size of the nuclei both in the laboratory frame when
each of the two nuclei is contracted up to a negligible small size, and
in the rest frame of one of the nuclei (target) when the second one
(projectile) passes through it as a  seemingly infinitely sharp shock
front. One cannot assign to the moving in the $x^+$ direction front a
finite width, neither in the $z$-, nor in the $x^-$-direction, without
a conflict with the special relativity. On the other hand, in the
framework of the wedge dynamics which operates with the boost states,
it is safe to consider the limit of the infinite momentum frame at the
end of the calculations. \footnote{This, however, leaves open the
question of what is detected in high-energy collision. The answer
crucially depends on what kind of the quantum mechanical ensemble is
involved in a particular measurement.}

\section{Scattering in wedge dynamics}
\label{sec:SN2}

\renewcommand{\theequation}{3.\arabic{equation}}
\setcounter{equation}{0}

The nuclei meet each other at the two-dimensional plane $t=z=0$, where
the first interaction take place. This interaction resolves the nuclei
constituents (e.g., the ``partons'', or ``color dipoles'') with the
boost $\nu\approx 0$, and excite the quantum states with the boost
$\nu\approx 0$. The wave functions of these states do not depend on the
rapidity coordinate $\eta$, and evenly fill in the interior of the
light wedge. At the same time, the two precursors, which are most
likely to be the fronts of the propagating gluon field, begin their way
in the light-like directions, $t\pm z=0$, thus creating the physical
boundaries of the light wedge, $\tau^2=t^2-z^2=0$. Passing through the
nuclei, the precursors resolve the elements with the finite boosts,
which are negative for the right-moving nucleus and positive for the
left-moving one, and initiate a transient process of interaction
between the nuclei. These interactions excite the quantum states with
positive and negative boosts, which depend on $\eta$ as
$~e^{i\nu\eta}$. In this way, the classical boosts, $\nu_{cl}=m_t
z_\ast$, are transformed into the quantum numbers of the wave functions
which have the period $2\pi/\nu$ in the $\eta$ direction, and occupy
the entire future domain of the point $t=z=0$. Before the collision,
the nuclei as a whole, are the coherent states of  QCD, and their
(color) coherence cannot be destroyed immediately. At $\tau\to +0$, the
resolved boost states have the same phases they had in the nuclei: the
decomposition of the nuclei in terms of the boost states is still a
coherent superposition. \footnote{The boundary condition
$A_\eta(\tau=0)=0$ imposed on the gauge fields in the wedge dynamics,
together with the gauge condition $A_\tau=0$, makes it impossible that
the fields of precursors immediately modify the phases (rotate the
color charges) along the light-like planes $x^+=0$ and $x^-=0$. This
property, which allows one to ``switch on'' the interaction between the
nuclei without an artificial color-changing ``shock wave," is in
contrast with the case of the null-plane dynamics with the gauges
$A^\pm=0$.}Furthermore, since the classical action of the states with
the finite boosts is large, even the resolved partons continue to move
along their rectilinear classical trajectories. The character of the
further evolution crucially depends on the subsequent interactions.
Below, we study the quasi-elastic forward scattering of the colored
quarks prepared and detected in the given boost states. This scattering
is mediated by the gluon field and results in the color exchange which
alone is capable of destroying the coherence of the nuclear wave
function.

The propagators of the gauge fields in wedge dynamics were studied in
\cite{gqm,geg}. In Appendix A, we review their properties with the
emphasis on the needs of the present study. The leading contribution
comes from the spatially local ``contact term'' of the longitudinal
part of the propagator. In order to give a flavor of its origin, we
have to emphasize, that we study the phenomenon where the finite charge
density is formed as a result of the interaction, and the proper fields
of the gradually created and yet delocalized charges physically overlap
with their sources. Thus, aiming at the dynamic picture, we have to
give the priority to the currents, expressing the charge density
$\rho(t)$ via the divergence of the current, $~\partial_t \rho
=-\nabla\cdot{\bf j}$, which eventually generates the contact term in
the propagator. The effect of the evolving charge density $\rho(t)$
becomes fully included into the Hamiltonian ${\cal H}_{int}={\bf
j}\cdot{\bf A}$, which is the only form compatible with the {\em
completely fixed} gauge $A^\tau=0$. This evolution of the color charge
density is the result of the interference between various partial
waves, and it is not connected with the motion of the physically
resolved point-like color charges. Without an interaction, these
partial waves would coherently sum and form the stable nuclei. Of those
interactions that take place when the nuclei intersect, the most
important are the ones which lead to the largest transition amplitudes.

An apparent complexity of the formulae in the wedge dynamics is caused
by the curvature of the hyper-surfaces of the constant $\tau$. The
hyper-surface $\tau=+0$ is the one where the initial data are naturally
set, and in has an infinite curvature. An explicit dependence of the
internal metric on $\tau$ makes the vector differential operators more
cumbersome and leads to an interplay between the longitudinal and
transverse fields.

\subsection{Choosing the observable}
\label{sec:Sb2.1}

Wedge dynamics deals only with the fields that emerged from
the localized collision of two macroscopic objects. This collision
is considered as a precise measurement of the partons coordinates
at the finite time moment $\tau\to +0$. Therefore, it is impossible to
pose a
formal scattering problem with the asymptotic initial states.
Instead, we take an approach based on the calculation of the
Heisenberg observable \cite{QFK,QGD},
\begin{eqnarray}
N (1',2')= \langle 1,2|\hat{n}(1')
(\hat{n}(2')-\delta_{1'2'})|1,2\rangle =
\langle 0| a_2 a_1
S^\dag a^\dag_{2'}a^\dag_{1'} a_{1'} a_{2'}
S a^\dag_1a^\dag_2 |0\rangle ~,
\label{eq:E2.1}\end{eqnarray}
which is the inclusively measured number of pairs of the final state
field excitations with quantum numbers $1'=(i'_1,k'_1)$ and
$2'=(i'_2,k'_2)$ ($k$ includes the transverse momentum and boost, $i$ -
color). This observable is evolved from the initial state of the two
interacting field  excitations with quantum numbers $1=(i_1,k_1)$ and
$2=(i_2,k_2)$. This quantity is closely related to the {\em total cross
section}. Indeed, assume that the measurementis an impulse process that
freezes decomposition of the colliding nuclei in terms of the
eigenfunctions of the corresponding operator. This decomposition
can become incoherent only due to real interaction, which will either
excite the new states, or just break the phase balance between the
initial ones. All this will contribute to the probability that the
initial state is altered, i.e., to the imaginary part of the forward
scattering amplitude. The color exchanges take place at the earliest
possible time $\tau_{min}\sim 1/\sqrt{s}$, and create a new color
composition which must eventually (with the probability one) evolve
into a new composition of hadrons. We emphasize, that a particular
choice of the basis of the interacting at $\tau>0$ boost states is
important only as long as we are interested in the rate at which the
color coherence is broken. The color transfer between the boost states
seems to be extremely intensive at the beginning of the collision. The
geometry of the collective modes which will be the actual final states
can be quite different \cite{fse}.

Expression (\ref{eq:E2.1}) is bi-linear with respect to the evolution
operator $S$ and thus it cannot be treated according to the Feynman
rules. For its evaluation on should use the so-called Schwinger-Keldysh
technique \cite{Keld} in the form adjusted for the calculation of
inclusive amplitudes \cite{QFK}. The evolution operator for the problem
of  evolution of the observable (\ref{eq:E2.1}) is of a usual form,
\begin{eqnarray}
S=T\exp\{i\int H_{int}(x)d^4x\}
\label{eq:E2.2}\end{eqnarray}
with the Hamiltonian
\begin{eqnarray}
H_{int}(x)= j^\mu(x) A_\mu (x)= j^\mu(x)[A^{[tr]}_\mu(x) +
\int dz D^{[long]}_{\mu\nu}(x,z) j^\nu (z)]~,
\label{eq:E2.3}\end{eqnarray}
where the second term in brackets is the longitudinal field
$A^{[long]}_\mu(x)$.  The propagator $D^{[long]}_{\mu\nu}(x,z)$
implicitly contains $\theta(x^0-z^0)$. For the sake of definiteness,
consider the fermion color current,
\begin{eqnarray}
j^\mu(x)_a=g~\overline{\Psi}_i(x)t^a_{ij}\gamma^\mu \Psi_j(x)~,
\label{eq:E2.4}\end{eqnarray}
and commute the final-state Fock operators with $S$ and
$S^\dag$ using the commutators,
\begin{eqnarray}
a_i(k) S-S a_i(k)=\int dz \overline{\psi}^{(+)}_k(z)
{\delta S\over \delta \overline{\Psi}_i(z)},\nonumber \\
S^\dag a^\dag_i(k) -a^\dag_i(k) S^\dag=
\int dz {\delta S^\dag \over \delta \Psi_i(z)}
\psi^{(+)}_k(z)~,
\label{eq:E2.5}\end{eqnarray}
In this equation, $\psi^{(+)}_k(z)$ is the one-particle wave
function from the decomposition of the field operator,
\begin{eqnarray}
\Psi_i(x)=\sum_k [a_i(k)\psi^{(+)}_k(x) +
b^\dag_i(k)\psi^{(-)}_k(x)]~.
\label{eq:E2.6}\end{eqnarray}

These commutations result in (disconnected pieces are omitted)
\begin{eqnarray}
N(1',2')= \int dx_1 dx_2 dy_1 dy_2
\overline{\psi}^{(+)}_{k'_2}(y_2)
\overline{\psi}^{(+)}_{k'_1}(y_1)\langle
0| a_2 a_1 {\delta^2 S^\dag \over \delta \Psi_{i'_2}(x_2)
\delta \Psi_{i'_1}(x_1)} \nonumber\\
\times~{\delta^2 S \over
\delta \overline{\Psi}_{i'_1}(y_1)
\delta \overline{\Psi}_{i'_2}(y_2)}
a^\dag_1a^\dag_2 |0\rangle \psi^{(+)}_{k'_2}(x_2)
\psi^{(+)}_{k'_1}(x_1)~.
\label{eq:E2.7}\end{eqnarray}
Here, the functional derivatives over $\Psi$ act from the left, and
the derivatives over $\overline{\Psi}$ act from the right. Next, we
compute the functional derivatives retaining the terms up
to the order $g^2$. This yields,
\begin{eqnarray}
N(1',2')=
g^4~\int dx_1dx_2dy_1dy_2 \overline{\psi}^{(+)}_{k'_2}(y_2)
\overline{\psi}^{(+)}_{k'_1}(y_1)~ \langle 0|~ a_{i_2}(k_2)
a_{i_1}(k_1) \nonumber\\ \times T^\dag
\big\{[- \overline{\Psi}_{l_2}(x_2)\gamma^\mu
\overline{\Psi}_{l_1}(x_1)\gamma^\nu
A^{[tr]a}_{\mu}(x_2)A^{[tr]b}_{\nu}(x_1) +
i\overline{\Psi}_{l_2}(x_2)\gamma^\mu D^{[long]ab}_{\mu\nu}(x_2,x_1)
\overline{\Psi}_{l_1}(x_1)\gamma^\nu  \nonumber\\
- i \overline{\Psi}_{l_1}(x_1)\gamma^\mu
D^{[long]ba}_{\mu\nu}(x_1,x_2)
\overline{\Psi}_{l_2}(x_2)\gamma^\nu]~\big\}~~
t^a_{l_2i'_2}t^b_{l_1i'_1}
t^{a'}_{i'_2j_2}t^{b'}_{i'_1j_1} \nonumber\\
\times T \big\{[ \Psi_{j_2}(y_2)\gamma^\mu
\Psi_{j_1}(y_1)\gamma^\nu
A^{[tr]a'}_{\mu}(y_2)A^{[tr]b'}_{\nu}(y_1) - i\gamma^\mu
\Psi_{j_2}(y_2) D^{[long]a'b'}_{\mu\nu}(y_2,y_1) \gamma^\nu
\Psi_{j_1}(y_1)  \nonumber\\
+ i \gamma^\mu \Psi_{j_1}(y_1)
D^{[long]b'a'}_{\mu\nu}(y_1,y_2)
\gamma^\nu \Psi_{j_2}(y_2)]~\big\}~
a^\dag_{i_1}(k_1)a^\dag_{i_2}(k_2)~ |0\rangle~
\psi^{(+)}_{k'_2}(x_2) \psi^{(+)}_{k'_1}(x_1)~.
\label{eq:E2.8}\end{eqnarray}

The calculations are accomplished as follows. The fermion operators are
contracted with the remaining Fock operators of the initial state,
producing the final-state wave functions, and making the final
adjustment of the color indices. This can be done in two ways, which
differ by a full interchange of the quantum numbers of the one-particle
initial states. The vacuum average of the products of the transverse
gluon field operators gives the transverse part of the $T$-ordered
propagator $D^{[00]}(y_2,y_1)$ and of the $T^\dag$- ordered
propagator $D^{[11]}(x_2,x_1)$. \footnote{ In this paper, we use the
Keldysh-Schwinger formalism \cite{Keld} in its modified form developed
earlier with the view of application to the inclusive and transient
processes. We employ the notation used in Refs.~\cite{QFK,QGD,tev}. The
indices  of the field correlators with the Keldysh contour ordering of
the
field operators (like $D_{[AB]}$) as well as the labels of their linear
combinations (like $D_{[ret]}$) are placed in square brackets.} The two
terms, with $D^{[long]}(y_2,y_1)$ and $D^{[long]}(y_1,y_2)$ cover two
complementary domains, $y^0_2>y^0_1$ and $y^0_2<y^0_1$, respectively.
Together, they form the longitudinal part of the $T$-ordered propagator
$D^{[00]}(y_2,y_1)$. Finally,
the transition probability can be cast in the form,
\begin{eqnarray}
N(1',2')= g^4~\bigg|~\int dx_1dx_2
[\overline{\psi}^{(+)}_{k_2}(x_2)\gamma^\mu
\psi^{(+)}_{k'_2}(x_2) \overline{\psi}^{(+)}_{k_1}(x_1)\gamma^\nu
\psi^{(+)}_{k'_1}(x_1)\nonumber\\ \times
D^{[00]}_{\mu\nu}(x_2,x_1)t^a_{i_2i'_2}t^a_{i_1i'_1}
-{\rm the~~same}(k_1,i_1\leftrightarrow k_2,i_2)]~\bigg|^2~.
\label{eq:E2.9}\end{eqnarray}

\subsection{Scattering of scalar quarks with the given boosts $\nu$.}
\label{sec:Sb3.2}

\begin{figure}[htb]
\begin{center}
\mbox{
\epsfig{file=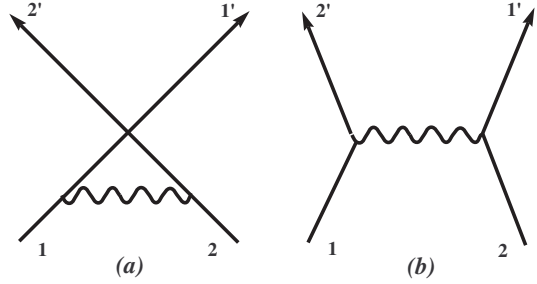,height=1.5in,bb=140 550 450 710} }
\end{center}
\caption{The forward (a) and backward (b) amplitudes of the $qq$
scattering. }
\label{fig:fig1}
\end{figure}

The observable  number of couples, $ N(1',2')$, can be rewritten by
introducing the full set of the intermediate states into the
Eq.~(\ref{eq:E2.1}),
\begin{eqnarray}
N (1',2')= \sum_{X}
\langle 0| a_2 a_1 S^\dag a^\dag_{2'}a^\dag_{1'} |X\rangle
\langle X|a_{1'} a_{2'} S a^\dag_1a^\dag_2 |0\rangle
=\sum_{X}|\langle X|a_{1'} a_{2'} S a^\dag_1a^\dag_2|0\rangle |^2~.
\label{eq:E2.1a}\end{eqnarray}
In the lowest order of the perturbation theory, there is no
additional emissions, and only the vacuum state
$ |X\rangle =|0\rangle$ contributes,
\begin{eqnarray}
N (1',2') = |M_{1,2\to 1',2'}|^2~.
\label{eq:E2.1b}\end{eqnarray}
In the lowest order, the inclusive transition probability
(\ref{eq:E2.1a}) is just the squared modulus of the matrix element
depicted on Fig.~\ref{fig:fig1}.

\subsubsection{Scattering amplitude}
\label{sec:Sbsb2.2.1}

Consider the matrix element of the scattering amplitude
\begin{eqnarray}
M_{1,2\to 1',2'}= g^2~\int dx_1dx_2 j^\mu_{k_2,k'_2}(x_1)
j^\nu_{k'_1,k'_1}(x_2)
D^{[00]}_{\mu\nu}(x_1,x_2)~.
\label{eq:E2.10}\end{eqnarray}
where we exchange the spinor quarks for the scalar ones, and
accordingly replace
$$ j^\mu_{k,k'}(x)= \overline{\psi}^{(+)}_{k} (x)
\gamma^\mu (x) \psi^{(+)}_{k'}(x) ~\to~ g^{\mu\nu}(x)
\overline{\psi}^{(+)}_{k}(x)~i
{\stackrel{\leftrightarrow}{\partial_\nu}} \psi^{(+)}_{k'}(x) ~,$$
using the states of scalar quarks with the quantum numbers
$k=({\vec k},\nu)$, transverse momentum and boost. In this case,
the wave functions are of the form,
\begin{eqnarray}
\psi^{(+)}_{{\vec k},\nu}(x)={e^{-\pi\nu/2}\over 2^{5/2}\pi }
H^{(2)}_{-i\nu} (m_{t}\tau)
e^{i\nu\eta +i{\vec k}{\vec r}}~,~~~~
\overline{\psi}^{(+)}_{{\vec k},\nu}(x)=
{e^{-\pi\nu/2}\over 2^{5/2}\pi } H^{(1)}_{i\nu}
(m_{t}\tau) e^{-i\nu\eta -i{\vec k}{\vec r}}~.
\label{eq:E2.11}\end{eqnarray}
Using the propagator in the mixed representation,
\begin{eqnarray}
D^{[00]}_{lm}(x_1,x_2)=\int{d\zeta d{\vec q}\over (2\pi)^3}
D^{[00]}_{lm}(\tau_1,\tau_2;\zeta, {\vec q})
e^{-i\zeta(\eta_1-\eta_2)-i{\vec q}( {\vec r}_1-{\vec r}_2)}~,
\label{eq:E2.12}\end{eqnarray}
and integrating over the spatial coordinates, we obtain
\begin{eqnarray}
M_{1,2\to 1',2'}= {g^2 \over 2^7 \pi}~ \delta(
\nu_1+\nu_2-\nu'_1-\nu'_2) \delta( {\vec k}_1+{\vec k}_2-{\vec
k}'_1-{\vec k}'_2) \nonumber\\ \times
\int_{0}^{\infty}\tau_1 d\tau_1
\int_{0}^{\infty}\tau_2 d\tau_2
H^{(1)}_{i\nu'_1} (m'_1\tau_1) H^{(2)}_{i\nu_1}
(m_1\tau_1)
H^{(1)}_{i\nu'_2} (m'_2\tau_2) H^{(2)}_{i\nu_2}
(m_2\tau_2)\nonumber\\
\times g^{ll}(\tau_1) g^{mm}(\tau_2) (k_1+k'_1)_l
(k_2+k'_2)_m D^{[00]}_{lm}(\tau_1,\tau_2;\zeta, {\vec q})~,
\label{eq:E2.13}\end{eqnarray}
where $\zeta=\nu_1-\nu'_1=\nu_2-\nu'_2$, ${\vec q}=
{\vec k}_1-{\vec k}'_1= {\vec k}_2-{\vec k}'_2$, and we introduced
3-vectors, $g^{ll}(\tau)p_l= (-{\vec p},
-\nu /\tau^2)~$, as well as a
short-hand notation, $m_i^2=m^2+{\vec k}_i^2$,
${m'_i}^2=m^2+{\vec k}_i^{'2}$.

Computing the transition amplitude (\ref{eq:E2.10}), we will be
interested in the states with the large boosts $|\nu | \gg 1$,
$\nu > m_{t}\tau $, $|\nu_1-\nu_2|\gg 1$ . In this case, the
asymptotic  of the Hankel functions reads as
\begin{eqnarray}
\pi e^{-\pi\nu/2} H^{(2)}_{-i\nu} (m_{t}\tau)=
\big[\pi e^{-\pi\nu/2}H^{(1)}_{i\nu} (m_{t}\tau)\big]^\ast
\approx {\sqrt{2\pi i}\over [m^2_{t}\tau^2+\nu^2]^{1/4}}~~
e^{-i\sqrt{m^2_{t}\tau^2+\nu^2}
+i\nu{\rm Arsh}(\nu/m_{t}\tau)}~.
\label{eq:E2.14}\end{eqnarray}
We have mentioned already, that in this limit, we have
$\psi^{(+)}_{{\vec k},\nu}(x) \propto \exp(iS_{cl})$,
where $S_{cl}$ is the classical action, found in Sec.~\ref{sec:SN1}.
In the limit of $|\nu_i| \gg m_i\tau $, we have
$$\nu{\rm Arsh} {\nu\over m \tau} = \nu\ln\bigg
[\sqrt{{\nu^2\over m^2 \tau^2}+1}+{\nu\over m \tau}\bigg]
\approx |\nu|\ln(2|\nu|)-\nu\ln(m\tau)~~,$$
and the product of the four Hankel functions in the
integrand of Eq.~(\ref{eq:E2.13}) becomes
\begin{eqnarray}
{4 (m_1)^{-i\nu_1} (m_2)^{-i \nu_2}
(m'_1)^{i \nu'_1}(m'_2)^{i \nu'_2} \over
\pi^2 |\nu_1\nu_2\nu'_1\nu'_2|^{1/2}}
e^{-i(|\nu_1|+|\nu_2|-|\nu'_1|-|\nu'_2|)} \nonumber\\
\times e^{-i[|\nu_1|\ln (2|\nu_1|)+
|\nu_2|\ln (2|\nu_2|)- |\nu'_1|\ln
(2|\nu'_1|)-|\nu'_2|\ln (2|\nu'_2|)]}
\bigg({\tau_1\over\tau_2}\bigg)^{-i\zeta}~~.
\label{eq:E2.14a}
\end{eqnarray}
The last factor here is the most significant for future
analysis. The rest is just the phase factor.

In what follows, we compute the leading term corresponding to
the contact part of the gluon propagator (see Eq.~(\ref{eq:A.26}) in
Appendix A),
\begin{eqnarray}
\big[ ~D^{[00]}_{\eta\eta}
(\tau_1,\tau_2;\eta,\vec{r})~ \big]_{contact}
=-{|\tau_1^2-\tau_2^2| \over 2} \delta(\eta)\delta(\vec{r}) ~.
\label{eq:E2.14b}
\end{eqnarray}
It is local in $\eta$ and $\vec{r}$, and the modulus accounts
for both terms with $D^{[long]}$ in Eq.(\ref{eq:E2.8}).
In this approximation, the matrix element (\ref{eq:E2.13}) becomes
\begin{eqnarray}
M_{1,2\to 1',2'}= {g^2 \over 2 (2\pi)^3}~
\delta(\nu_1+\nu_2-\nu'_1-\nu'_2)
\delta( {\vec k}_1+{\vec k}_2-{\vec k}'_1-{\vec k}'_2)
{(\nu_1+\nu'_1)(\nu_2+\nu'_2) \over 4 |\nu_1\nu_2\nu'_1\nu'_2|^{1/2}}
~e^{i\alpha}~ I ~,
\label{eq:E2.13a}\end{eqnarray}
where $\alpha$ is an inessential real phase. In the approximation
given by the equation (\ref{eq:E2.14a}) it absorbs all the
dependence on the transverse momenta. Now, it remains  to
compute the integral
\begin{eqnarray}
I=\int_{0}^{\infty}\tau_1 d\tau_1\int_{0}^{\infty}\tau_2 d\tau_2
g^{\eta\eta}(\tau_1) g^{\eta\eta}(\tau_2) \big[
~D^{[00]}_{\eta\eta}(\tau_1,\tau_2)~ \big]_{contact}
\bigg({\tau_1\over\tau_2}\bigg)^{-i\zeta} \nonumber \\
=\int_{\tau_0}^{T}d\tau_1\int_{\tau_0}^{T}d\tau_2
{|\tau_1^2-\tau_2^2| \over 2\tau_1\tau_2}
\bigg({\tau_1\over\tau_2}\bigg)^{-i\zeta}~,
\label{eq:E2.15}\end{eqnarray}
where the cutoffs are introduced in order to
isolate the possible singular behavior. Computation is
straightforward,
\begin{eqnarray}
2I=\int_{\tau_0}^{T}\tau d\tau \int_{\tau_0/\tau}^{1}dx
[x^{i\zeta~}+ x^{-i\zeta~}]\bigg({1\over x}-x \bigg)
\hspace{5.5cm}\nonumber\\
={T^2\over \zeta^2+4}\bigg\{4\bigg(1+{\tau_0^2\over T^2}\bigg)~
{\sin[\zeta\ln(T/\tau_0)]\over \zeta}- 2\bigg(1-{\tau_0^2\over
T^2}\bigg) \big(1+\cos[\zeta\ln(T/\tau_0)]\big)~\bigg\}~,
\label{eq:E2.16}\end{eqnarray}
The cutoffs $\tau_{min}=\tau_0$ and $\tau_{max}=T$ in this formula are
the external physical input. Making a choice for $\tau_0$ and $T$ it is
useful to keep in mind that the interaction (\ref{eq:E2.14b}) is due to
a non-stationary part of the longitudinal (Coulomb) field of the
charges resolved at $\tau=0$. Similar cutoffs are needed in a
stationary part. A proper choice leads to the Coulomb logarithm in the
collision term in the QED plasma, and we follow this example.

The only field $A_\eta$ that contributes the contact term
(\ref{eq:E2.14b}) vanishes at $\tau=0$, and its effect on the resolved
at $\tau=0$ charges cannot be instantaneous. Therefore, the lower limit
$\tau_0$ is related to the earliest time when the boost states
belonging to the incoming nuclei are resolved by means of the strong
interaction. Practically, this is the time which it takes
two nuclei to  completely overlap. Therefore, this minimal time is
defined by the velocities of the incoming nuclei in the laboratory
frame, $\tau_{0}\sim 1/\sqrt{s}$. At this time, the stationary phase of
partial waves (\ref{eq:E1.20}), $\omega^{(+)}_{{\vec k},\theta}(x)$,
corresponding to the particles with the given rapidities $\theta$, are
stretched over the widest rapidity interval $\Delta\eta\sim
2\ln(\sqrt{s}/m_{\perp})$. This estimate coincides with the well known
kinematically allowed width $2Y$ of the rapidity plateau, $2Y\approx
log(s/m^2_{char})$.(See Ref \cite{gqm} for the further details.)

The upper limit $T$  has to be set because at some time $\tau_{max}$
the picture of the independent collisions breaks up. The ``final''
state fields are not emitted into the free space any more (which
affects even the QCD evolution equations \cite{tev}). Therefore, $T$
corresponds to the time when  subsequent interactions begin to erase
the memory about the origin of the boost states from the compact
nuclei. By this time, the system must develop collective interactions
which result in the effective masses of the plasmon-like modes in
a dense medium. It is clear that these masses can emerge only gradually
\cite{tev,gqm}. An attempt to  evaluate this gradual process in the
scope of wedge dynamics has been undertaken in Refs. \cite{tev,fse}.
This calculation relies on the following physical mechanism:
The low--$p_t$ mode of the radiation field acquires a finite effective
mass as a result of its forward scattering on the strongly localized
(and formed earlier) particles with $q_t\gg p_t$.
Regardless of what the exact value of this ``screening mass'' $\mu_D$
is, it seems reasonable to take $T\sim 1/\mu_D$, which is consistent
with the semi-classical approximation, $T\mu_D\ll\nu$.

The two limits of the Eq.(\ref{eq:E2.16}) are of special interest.
Let $\sqrt{s}\to \infty$, while $\zeta$ is kept finite. Then
\begin{eqnarray}
I \sim {T^2\over \zeta^2+4}\bigg\{2\pi\delta( \zeta)-1~\bigg\}~,
\label{eq:E2.17}\end{eqnarray}
the amplitude is strongly confined near the forward region and
the corresponding cross section diverges.

Next, let us consider the physical limit of the forward scattering,
$\zeta\to 0$, while keeping $\sqrt{s}$ finite. In this case, we have
\begin{eqnarray}
I \sim T^2 [\ln(T\sqrt{s})-1]~,
\label{eq:E2.18}\end{eqnarray}
the inclusive amplitude is proportional to the maximal width of the
rapidity plateau, $Y\propto \ln(\sqrt{s})$, which is the only geometric
factor that can accompany the contact interaction  (\ref{eq:E2.14b}).
Its square naturally sets the upper bound for the scattering
probability.

The second term in the forward scattering amplitude
(\ref{eq:E2.9}),which corresponds to the complete exchange of the two
initial states (backward scattering), is obviously small. Indeed, this
case corresponds to $\nu_1\approx\nu'_2$, and $\nu_2\approx\nu'_1$. In
this case, $$|\zeta|=|\nu_1-\nu'_1| \approx |\nu_1-\nu_2| \gg 1~,$$ and
the function (\ref{eq:E2.16}) is small.

\subsubsection{Scattering probability}
\label{sec:Sbsb3.2.2}

Since we consider the processes which develop in the course of a
single collision, the notion of the cross section is not well
defined. In order to deal with the quantity which is as close as
possible to the standard cross section, let us introduce the
``normalization volume''  $\Omega=\pi R^2 Y$, the product of
the transverse area and the length of the rapidity interval over
which the nuclei become expanded by the first measurement of the
collision coordinates. The wave functions of all states begin to
occupy this volume when the two nuclei have completely
overlapped, i.e., by the time $\tau_{min}\sim 1/\sqrt{s}$.  The
wave functions $\psi_{{\vec k},\nu}$, given by Eq.
(\ref{eq:E2.11}), in the matrix element (\ref{eq:E2.9}) thus
acquire an additional factor $(2\pi)^{3/2}\Omega^{-1/2}$. The
quantity $\rho=\Omega^{-1}$ will play the same role as the
flux factor $j=1/ST=v_{rel}/V$ in the case of the standard $2\to n$
scattering (see, e.g. Ref. \cite{Bjorken}). Multiplying the squared
modulus of the matrix element (\ref{eq:E2.13a}) by the densities of
the final states, $\Omega d^2{\vec k}' d\nu'/ (2\pi)^{3/2}$, and
replacing one of the delta functions by $\Omega/(2\pi)^3$,we arrive at
the
differential inclusive probability,
\begin{eqnarray}
d w ~= ~{ \delta(\nu_1+\nu_2-\nu'_1-\nu'_2)
\delta({\vec k}_1+{\vec k}_2-{\vec k}'_1-{\vec k}'_2)
\over \Omega}\nonumber\\
\times~{\alpha_s^2 \over 2\pi}~
{(\nu_1+\nu'_1)^2(\nu_2+\nu'_2)^2
\over 16|\nu_1\nu_2\nu'_1\nu'_2|}~ I^2
d^2{\vec k'}_1 d\nu'_1 d^2{\vec k'}_2 d\nu'_2~.
\label{eq:E2.13b}\end{eqnarray}
Dividing $d w$ by the density $\rho=\Omega^{-1}$, we obtain the
closest analog of the cross section which can be introduced in order
to characterize a {\em single event},
\begin{eqnarray}
d \sigma_1 ~= ~ \delta(\nu_1+\nu_2-\nu'_1-\nu'_2)
\delta({\vec k}_1+{\vec k}_2-{\vec k}'_1-{\vec k}'_2) \nonumber\\
\times~{\alpha_s^2 \over 2\pi}~
{(\nu_1+\nu'_1)^2(\nu_2+\nu'_2)^2
\over 16|\nu_1\nu_2\nu'_1\nu'_2|}~ I^2
d^2{\vec k'}_1 d\nu'_1 d^2{\vec k'}_2 d\nu'_2~.
\label{eq:E2.13c}\end{eqnarray}
Since $I^2$ has the dimension $ [length]^4$, the
quantity $\sigma_1$ also has the dimension of area.
The upper limit $\tau_{max}=T$ in
Eqs.~(\ref{eq:E2.15})--(\ref{eq:E2.18})
can be estimated from the condition $\tau \mu_D\approx 1\ll\nu$,
and is related to the formation of the (final) states as they are
detected by the subsequent interactions at
the later period of the evolution. In the limit of a nearly
forward scattering, and integrating  $d^2{\vec k'}_2 d\nu'_2$
with the aid of the delta-functions, we arrive at
\begin{eqnarray}
{d \sigma_1\over d^2{\vec k}'_t d\zeta}~=
~{\alpha_s^2 \over 2\pi}~~{2\over 9}~~
{(2\nu_1+\zeta)^2(2\nu_2-\zeta)^2
\over 16|\nu_1\nu_2(\nu_1+\zeta)(\nu_2-\zeta)|}~
{ 1 \over \mu_D^4}~\bigg({2\over \zeta^2+4}\bigg)^2 \nonumber\\~
\times \bigg[ {\sin[\zeta \ln(\sqrt{s}/ \mu_D)]\over \zeta} -
{1+\cos[\zeta \ln(\sqrt{s}/ \mu_D)]\over 2}\bigg]^2 ~.
\label{eq:E2.13d}\end{eqnarray}
In the limit of the forward scattering it becomes
\begin{eqnarray}
\bigg[{d \sigma_1\over d^2{\vec q_t} d\zeta}\bigg]_{\zeta\to 0}~=
~{\alpha_s^2 \over 8\pi}~~{2\over 9}~~
~{ 1 \over \mu_D^4}~ \ln^2{\sqrt{s}\over \mu_D}~,
\label{eq:E2.13e}\end{eqnarray}
where ${\vec q}_t\approx {\vec k'}_t$ is considered as the
transverse momentum transfer. Our basic approximation
implies that this transfer is small, $q_t<\mu_D$. The  color trace
$$ {2\over 9} = {1\over 3}\cdot {1\over 3}\cdot \bigg({6\over
4}+{2\over 4}\bigg)$$ accounts for the processes with and without
color transfer.

\section{Summary}
\label{sec:SN3}

The main result of this note is given by the Eq.~(\ref{eq:E2.13e}). The
logarithmic character of the answer (the color-changing amplitude
$\propto \alpha_s \ln(\tau_{min}/\tau_{max})\approx \alpha_s
\ln(\sqrt{s}/\mu_D)~)~$ is due to the dimensionless-ness of the
rapidity and the boost variables, rather than to the Coulomb nature of
the interaction. This answer indicates, that we may expect a massive
breakdown of the color balance  in the colliding nuclei at
the earliest time $\tau \sim 1/\sqrt{s}$. The rate at which the
intensity of this breakdown grows with the energy is proportional to
$\ln^2s$.

The key assumption that led to this result is the existence of a sharp
boundary of the colliding nuclei. If this assumption is not correct,
then there is no any reason to consider the problem of the nuclear
collision in the framework of wedge dynamics, and the whole picture of
the collision will look differently. This would also undermine
alternative approaches to the problem, like the McLerran-Venugopalan
model \cite{McLerran1,Kovchegov}. An immediate logical consequence of
the finite size is the absence, inside the stable nuclei, of the finite
color charge density, which could significantly fluctuate and produce
the long-range fields. Only under this assumption could we safely
discard the static component of the gauge field which would correspond
to the finite charge density at $\tau=0$ and to consider the creation
of color charges in the course of the nuclear collision as a transient
process. The currents in rapidity direction, which we relied upon in
our calculations, appear as a result of the phase shifts in the system
of delocalized fields (and thus propagating with the phase velocity),
rather than due to the motion of the resolved point-like charges.

The wedge dynamics was conceived as a tool which is adequate to the
earliest stage of the collision, where the initial color coherence
becomes broken. It is not applicable to the $ep-$DIS, where the
electron probes the long range electromagnetic fluctuations in the
proton \cite{QGD,tev}. In its turn, the evolution equations which
describe QCD fluctuations that accompany $ep$-DIS do not seem to be
relevant to the collisions of the finite-size nuclei. The primary
breakdown of color coherence in a nuclear collision (in terms of the
states of the wedge dynamics, it is indeed the earliest process) must
result in color radiation which can exist only for a short period (in
proper time), only before the fields begin to build up the collective
modes of the expanding continuous media. Only these collective effects
can bring in the scale ($\mu_D$) to the entire process and serve as a
feed-back that limit the intensity of the primary emissions
\cite{tev,fse}. Later on, the dynamics of the process must become local
on this scale.

The transient process of the plasma formation will come to its
saturation at the moment when the growing with time (and density)
effective masses of the collective modes begin to screen all emission,
from the evolving sources, at the scales below the one given by the
dynamically generated effective masses \cite{tev,fse}. Being unable to
radiate, these sources must pass through and form the receding nuclear
remnants.  Thus, it is likely that the total energy of the collision is
responsible only for the time-scale of the initial interaction and the
full width of the rapidity plateau, while the parameters of
the final state in the central rapidity region are universal and
independent of the initial energy (above a certain threshold).
Eventually, the total energy of the colliding nuclei is shared by the
newly born matter and these receding remnants. It is not clear yet if
the quark-gluon matter will have time to sufficiently thermalize and be
described by a single parameter, the temperature. However, it seems
unavoidable that the entropy created at the earliest moments
must result in the {\em pressure}, which is the first thing we shall
try to theoretically estimate. A success at this point will much
simplify the whole scenario by allowing to incorporate the hydrodynamic
picture from a sufficiently early proper time.

Our preliminary estimates show that the boost states of  wedge dynamics
do not effectively scatter with large transverse momentum transfer.
Further analysis is necessary to verify this estimate, which (being
correct) could explain the absence of high-$p_t$ jets observed in the
first available RHIC data. The jets are not strongly quenched, they
well can be absent at all! Does this mean that perturbative QCD is
totally unrelated to the ultrarelativistic heavy ion collisions? We do
not think so. It just has to be used in a different way than in {\em
ep}-DIS or {\em pp}-collisions. The major source of this difference has
been first outlined in Ref.~\cite{tev}: in nuclear collisions, the
final states that saturate the unitary cut in the ladders that
correspond to QCD evolution equations cannot be saturated by
quark and gluon states in free space.  In this paper, we point to the
fact that the initial states can be different from the free massless
wee partons with the given light-cone momenta. They well can be the
boost states of valence quarks which are explicitly confined inside the
finite-size nuclei before the collision.

\vspace{1cm}

\noindent {\bf ACKNOWLEDGMENTS}

The author is indebted to Edward Shuryak for continuous support, many
conversations and constructive critical remarks at various stages of
this work.
Discussions with Pawel Danielewich, Scott Pratt, Eugene Surdutovich and
Vladimir Zelevinsky were very helpful.

\section{Appendix A. The gluon propagator.}
\label{sec:SNA}

\renewcommand{\theequation}{A.\arabic{equation}}
\setcounter{equation}{0}

In order to study the interaction of the two charged states with
the given boosts $\nu$, one needs to have an explicit form of the
gauge field propagator. Particularly, since the boosts are
additive and obey the conservation law, one need to know what the
quanta of radiation that carry the boost quantum numbers are. It
is also necessary to know the form of the proper (longitudinal)
fields produced by the charged particles. In this section, we
present a detail analysis of the gluon propagator in wedge
dynamics, which has been derived in Ref.~\cite{geg}. The main
purpose is to carefully trace the origin of the new contact term.
At first glance, it may look abnormal since it neither shows up
in the field of a moving static charge, nor has it any
properties associated with the propagation. We want to show that
all Coulomb-type terms still exist in the propagator. They are
somewhat modified, in a way which one could expect on purely
physical grounds. Namely, the Coulomb fields vanish outside the
future domain of the point where the charge was created. Our
analysis indicates that the other parts of the propagator cannot
hide anything similar to the exclusive contact part which is solely
responsible for the final result, Eq.~(\ref{eq:E2.13e}), of this paper.

\subsection{The field of a static source}

The field of a static source in wedge dynamics is found \cite{geg}
when one solves the linearized (Maxwell) equations of motion
without the external current, imposing the gauge condition $A^\tau
=0$. An additional boundary condition, which allows one to fix the
gauge completely, is $A_\eta(\tau=0)=0$. In fact, this condition
brings nothing new, since the hypersurface $\tau=0$ is light-like,
and the $\tau$-  and $\eta$-directions are degenerate there. In
this way, one finds {\em three} modes, of which two,
$V^{(TE)}_{\nu{\vec k}}(x)$ and  $ V^{(TM)}_{\nu{\vec k}}(x)$ are
the transverse fields. The modes $V^{(TE)}$ and $ V^{(TM)}$ are
normalized according to a usual definition of the scalar product
in the functional space of the solutions of the Maxwell equations,
\begin{eqnarray}
(V,W)=\int_{-\infty}^{\infty}d\eta\int d^2{\vec r}
\tau {\sf g}^{ik}
V^{*}_{i} i{\stackrel{\leftrightarrow}{\partial}}_{\tau} W_k ~,
\label{eq:A.1}\end{eqnarray}
and satisfy the Gauss law without the charge. The third mode,
$V^{(stat)}$ has zero norm, and its definition is accomplished with
the aid of Gauss law with the {\em static} source. [In the absence
of any currents, the source can be only static.] The electric
and magnetic fields of this mode are
\begin{equation}
E_l^{[stat]}(\tau,{\vec r},\eta)=
\int {d\nu d^2{\vec k}\over (2\pi)^3}
{e^{i\nu\eta +i{\vec k}{\vec r}}
\over i k_{t}^2}
\left[
\begin{array}{c}
k_r \tau^{-1}  s_{1,i\nu}(k_{t}\tau) \\
\nu k_{t}^2 \tau s_{-1,i\nu}(k_{t}\tau)
\end{array}
\right]_l
\rho({\vec k},\nu)~.
\label{eq:A.2}
\end{equation}
\begin{eqnarray}
B^{[stat]}_{l}(\tau,{\vec r},\eta)=
\int {d\nu d^2{\vec k}\over (2\pi)^3}
{e^{i\nu\eta +i{\vec k}{\vec r}}\over k_{t}^2}
\left[ \begin{array}{c}
k_y   \\ - k_x\\ 0\end{array} \right]_l
\nu \dot{s}_{-1,i\nu}(k_{t}\tau)
\rho({\vec k},\nu)~~,
\label{eq:A.3}\end{eqnarray}
where $s_{m,i\nu}(x)$ is the Lommel function, a solution of the
inhomogeneous Bessel equation with $x^{m-1}$ as the external source,
$$ f''+{1\over x}f'+\bigg(1+{\nu^2\over x^2}\bigg)f=x^{m-1}~.$$

There exists an extremely important relation between the
two Lommel functions,\footnote{ It is useful to keep in mind the
integral
representation
\begin{eqnarray}
s_{1,i\nu}(k_{t}\tau)=
1-{\nu\over\sinh\pi\nu}\int_0^\pi\cos(k_{t}\tau\sin\phi)
\cosh\nu\phi d\phi~~,
\label{eq:A.19}\end{eqnarray}
which indicates that the functions $s_{1,i\nu}(x)$ and $\nu^2
s_{-1,i\nu}(x)$ are regular at $\nu=0$.}
\begin{eqnarray}
s_{1,i\nu}(k_{t}\tau) + \nu^2 s_{-1,i\nu}(k_{t}\tau) = 1~~.
\label{eq:A.8}\end{eqnarray}
First of all, it is necessary in order to verify that the electric
 field of a static charge distribution, (\ref{eq:A.2}), indeed
 satisfies Gauss law,\footnote{In terms of the physical components,
${\cal E}^m=\sqrt{-g}g^{mn}E_n=\sqrt{-g}g^{mn}\partial_\tau A_n ,$
the Coulomb law reads exactly as in Cartesian coordinates,
$\partial_m {\cal E}^m = \rho .$}
\begin{eqnarray}
{1\over\tau}\partial_\eta E_\eta +\tau\partial_r E_r =
\tau j^\tau=\rho~.
\label{eq:A.9}\end{eqnarray}
Second, it is precisely the unit on the right side of
Eq.~(\ref{eq:A.8}) which will give rise to the contact term in
the full propagator.

The Fourier component of the vector potential of the static field is
\begin{eqnarray}
A^{[stat]}_{l}({\vec k},\nu;\tau)= { \rho({\vec k},\nu)
\over (2\pi)^3 i k_{t}^2}
\left[ \begin{array}{c}
k_r  Q_{-1,i\nu}(k_{t}\tau) \\
\nu Q_{1,i\nu}(k_{t}\tau)
\end{array} \right]_l~~,
\label{eq:A.4}\end{eqnarray}
where we introduced the functions
 $$Q_{m,i\nu}(x)=\int_0^x x^m~s_{-m,i\nu}(x)~dx~.$$
In spite of an unusual appearance, this is nothing else but
Coulomb's law in the framework of wedge dynamics.
In order to see this explicitly, let us consider the system
of point-like charges located at the points  ${\vec r}_i $
in the transverse plane and moving with rapidities $\theta_i$,
\begin{eqnarray}
\rho=\tau j_\tau=\sum_i q_i~\delta(\eta-\theta_i)
\delta({\vec r}- {\vec r}_i).
\label{eq:A.5}\end{eqnarray}
 For a single charge, the explicit form of the electric field
components is
\begin{eqnarray}
E_l^{[stat]}(\tau,{\vec r},\eta)=
{q \over 4\pi}
\left[ \begin{array}{c}
{\vec r} \cosh(\eta-\theta) \\
\tau^2 \sinh(\eta-\theta)   \end{array} \right]_l
{\theta(\tau -r_{t})\over R^{3} }  +
\left[ \begin{array}{c}
{\vec r}/r_t^2 \\  \tanh(\eta-\theta)   \end{array} \right]_l
\delta(\tau -r_{t})~~,
\label{eq:A.6}\end{eqnarray}
where  $~R=[{\vec r}^2+\tau^2\sinh^2(\eta-\theta)]^{1/2}~,$ is the
distance between the points $({\vec 0},\theta)$ and $({\vec r},\eta)$
in the internal geometry of the surface $\tau=const$. On can obtain
the first term in this formula taking the usual  (gauge-independent)
expression for the electric field of the moving charge, transforming
it to the new coordinates, and multiplying it by the $\theta(\tau -
r_{t})$, which eliminates the field outside the light cone of the point
where the charge had emerged. The second term (with the light cone
delta-function) corresponds to the wave front that accompanies
the process of the charge creation at $\tau=0$.\footnote{This is not a
true radiation. The real Coulomb mode $ A_l^{[stat]}$ is {\em
orthogonal} to the complex propagating modes $V_l^{(TE)}$ and
$V_l^{(TM)}$.}

Since the electric field is $E_l=\partial_\tau A_l$, the vector
potential is recovered by means of integration,
\begin{eqnarray}
A_l(\tau)=\int_0^\tau E_l(\tau') d\tau'~ \to~\int_{r_t}^\tau
E_l(\tau') d\tau'.
\label{eq:A.7}\end{eqnarray}
Now, when $r_t$ is taken as the actual lower limit, the result of
the integration explicitly coincides with the Fourier transform of
Eq.~(\ref{eq:A.4}). The Fourier transform of the Lommel functions
appears to be discontinuous in an exactly relativistic way (the
details of its calculation are in the next section).

One may ask how the Coulomb mode could be found from  Maxwell's
equations of motion which do not include  Gauss' law. The answer
is simple and natural: the Coulomb field outside the static charge
distribution must satisfy the equations of motion for a free
field.

There are two surprises connected with the static solutions of the
wedge dynamics. First, {\em the source is static if  it expands in
such a way that its physical component} $~{\cal J}^\tau=\tau
j^\tau(\tau,\eta, {\vec r})~$ {\em does not depend on} $~\tau$.
Indeed, the charge conservation has its physical form,
$\partial_\mu {\cal J}^\mu = 0 ,$ only in terms of the physical
components ${\cal J}^\mu= \sqrt{-g}g^{\mu\nu} j_\nu $ of the
electric current. The second surprise is the light-cone boundary
of the static field in Eq.~(\ref{eq:A.6}).

Finally, let us consider the conservation of the charge of a
fundamental field in full QCD. Now, the equation of charge
conservation reads as
\begin{eqnarray}
\partial_\mu {\cal J}_a^\mu +
{\rm g}f^{abc}A^b_\mu{\cal J}_c^\mu =0~.
\label{eq:A.10}\end{eqnarray}
Let only the $j_a^\tau$ component of the current differs from zero.
Then for the charge $Q^a=\int \tau d\eta d^2{\vec r}j^\tau$, we have
\begin{eqnarray}
\partial_\tau Q_a + {\rm g}f^{abc}A^b_\tau Q_c =0~.
\label{eq:A.11}\end{eqnarray}
Since the gauge condition is $A_\tau =0$, we conclude that $Q_a
=const$. In the framework of wedge dynamics, the notion of static
charge is well defined even if the individual charges move with
respect to each other (in a specific way). Similar result can be
obtained in the system with the Hamiltonian time $t=x^0$ with the
gauge condition $A^0=0$. If all (color) charges are at rest, their
proper static field does not ``rotate'' their color. However, this
will not be the case if we chose a different gauge condition, e.g.
$A^3=0$ or ${\rm div} {\bf A}=0$, which would require that
$A^0\neq 0$.

Finally, it is easy to understand, that since the proper gluon field of
the static fundamental color charge does not affect  the charge itself,
this gluon field cannot be a carrier of the color charge. This fully
agrees with the fact, that the norm of the Coulomb mode equals zero,
because its field is real (contrary to the complex fields of the
transverse modes that represent gluons). An additional reason to pay
special attention to the static field configuration is that the field
corresponding to the charge density $\rho(\tau=0)$ is an isolated
exceptional static field. It was necessary to describe it in  details
in order to have a reference point for a more involved analysis of the
fields created by the charged currents.

\subsection{The full longitudinal field}

The gluon propagator, which we review and analyze in some details
below, was found as a (retarded) response function between the
potential and the current for the linearized (Maxwell) equations of
motion. The potential is represented as a sum of three terms,
$$A=A^{[tr]}+A^{[L]}+A^{[inst]}=A^{[tr]}+A^{[long]}.$$
The second and the third terms constitute the longitudinal (in a sense
of the Gauss' law) field. The goal of this somewhat technical analysis
is to demonstrate that the longitudinal part of this propagator indeed
includes a new contact term. At the same time we want to show, that the
standard Coulomb fields are still present in the propagator, almost
unchanged and are modified only by the relativistic causal boundaries
which one would expect to appear for the fields of the emerging
charges.

The transverse part of the retarded propagator is trivial. It is built
from the partial solutions of the homogeneous wave equations,
\begin{eqnarray}
A^{[tr]}_{l}(x_1)=\int d^4 x_2 \theta(\tau_1-\tau_2)
\Delta^{(tr)}_{lm}(x_1,x_2) j^m(x_2)~~.
\label{eq:A.12}\end{eqnarray} where
\begin{eqnarray}
\Delta^{(tr)}_{lm}(x,y)= -i\int_{-\infty}^{\infty} d\nu\int
d^2{\vec k}\sum_{\lambda=TE,TM}
[V^{(\lambda)}_{\nu {\vec k};l}(x)
V^{(\lambda)\ast }_{\nu {\vec k};m}(y)- V^{(\lambda)\ast}_{\nu
{\vec k};l}(x) V^{(\lambda) }_{\nu {\vec k};m}(y)]~~,
\label{eq:A.13}\end{eqnarray}
which can be easily recognized as
the Riemann function of the original homogeneous hyperbolic
system. The Riemann function solves the boundary value problem for
the evolution of the  free radiation field.  It is obtained
immediately as a bilinear expansion over the full
set of solutions  of the homogeneous system.

The name of the instantaneous part is motivated by its explicit
form,
\begin{eqnarray}
A^{[inst]}_{l}({\vec k},\nu;\tau_1)= { \rho({\vec k},\nu,\tau_1)
\over (2\pi)^3 i k_{t}^2} \left[ \begin{array}{c}
k_r  Q_{-1,i\nu}(k_{t}\tau_1) \\ \nu Q_{1,i\nu}(k_{t}\tau_1)
\end{array} \right]_l~~,
\label{eq:A.14}\end{eqnarray}
the {\em potential} $A^{[inst]}_{l}$ is simultaneous with the
charge density $\rho=\tau j_\tau$. Formally, it can be obtained
by adding the time dependence to the charge density in the
expression for the static potential (\ref{eq:A.4}).
However, this form is  inconvenient as long as we have to use
$A_\mu j^\mu=A_l j^l$ as the basic form of the interaction
Hamiltonian. Therefore, we have to eliminate the
charge density $\rho$ completely, and  replace it by the
spatial components  $j^n$ of the current.  The replacement
follows an evident prescription,
\begin{eqnarray}
\rho(\tau_1, \nu,{\vec k})-\rho(0,\nu,{\vec k}) =
\int_{0}^{\tau_1}d\tau_2 {\partial\rho \over\partial\tau_2} = -i
\int_{0}^{\tau_1}\tau_2 d\tau_2
[k_s j^s(\tau_2,\nu,{\vec k})+ \nu j^\eta(\tau_2,\nu,{\vec k})]~~.
\label{eq:A.15}\end{eqnarray}
The effect of the initial charge density $ \rho_0=\rho(0,\nu,{\vec k})$
would correspond to the clearly visible static pattern in the
longitudinal part of the field. In the framework of  perturbative QCD,
this pattern is not active, since it cannot transmit the color charge.
Furthermore, as we have argued previously, in nuclear collisions, the
initial density of the color charges at $\tau=0$ is zero. This leads to
\begin{eqnarray}
A^{[inst]}_{l}(\tau_1,\nu,{\vec k})= -\int_{0}^{\tau_1} {\tau_2
d\tau_2 \over  (2\pi)^3 k_{t}^2} \left[ \begin{array}{c}
k_r  Q_{-1,i\nu}(k_{t}\tau_1) \\
\nu Q_{1,i\nu}(k_{t}\tau_1)
\end{array} \right]_l
\left[ \begin{array}{c} k_s  \\
\nu  \end{array} \right]_m j^m(\tau_2,\nu,{\vec k})~~.
\label{eq:A.16}\end{eqnarray}
In this form, the three remaining (in
the gauge $A^\tau=0$) spatial components $A_l$ of the vector
potential are expressed via the spatial components of the current.

The dynamical  longitudinal field $A^{(L)}$ is of the form:
\begin{eqnarray}
A^{[L]}_{l}(\tau_1,\nu,{\vec k})=\int_{0}^{\tau_1} {\tau_2 d\tau_2
\over (2\pi)^3 k_{t}^2}
\left[ \begin{array}{c}
k_r  \\
\nu  \end{array} \right]_l
\left[ \begin{array}{c}
k_s  Q_{-1,i\nu}(k_{t}\tau_2) \\
\nu Q_{1,i\nu}(k_{t}\tau_2)
\end{array} \right]_m  j^m(\tau_2,\nu,{\vec k})~~.
\label{eq:A.17}\end{eqnarray}
It also does not allow for the bilinear expansion with two temporal
arguments. Its electric and magnetic field is simultaneous with the
current $j^m$ also. In what follows immediately, we intend to single
out the contact part of the propagator, which shows up only in the
$D_{\eta\eta}$ component and connects $A_\eta$ with $j_\eta$.

In order to set the stage, it is instructive to  start with the
electric and magnetic  fields of these two modes,
$E_m=\stackrel{\bullet}{A}_m$,
$ {\cal E}^m=\sqrt{-{\rm g}}{\rm g}^{mn} \stackrel{\bullet}{A}_n$
and ${\cal B}^m=-(2\sqrt{-{\rm g}})^{-1} e^{mln}F_{ln}$.
Since the potential $A^{[L]}_{l}$ is  the three-dimensional gradient,
we immediately see that $ {\cal B}^{[L]}_{l}=0$.
[Note, that $A^{[L]}_{l}$ is the gradient of a {\em time-dependent}
function, and thus is not a pure gauge.] Starting from the
expression for $A^{[inst]}$, and using the relation \cite{geg},
\begin{eqnarray}
Q_{-1,i\nu}(k_{t} \tau)-Q_{1,i\nu}(k_{t} \tau)= -{\tau \over
\nu^2} {\partial\over\partial\tau} s_{1,i\nu}(k_{t}\tau)= \tau
{\partial\over\partial\tau} s_{-1,i\nu}(k_{t}\tau)~~,
\label{eq:A.18}\end{eqnarray}
we obtain by a straightforward calculation that
\begin{eqnarray}
{\cal B}^{[inst]}_{l}(\tau,\nu,{\vec k})=
\int_{0}^{\tau} {\tau_2
d\tau_2\nu \over  (2\pi)^3 i k_{t}^2} \left[ \begin{array}{c}
k_y   \\ - k_x\\ 0\end{array} \right]_l
\left[ \begin{array}{c}  k_x \\ k_y  \\
\nu  \end{array} \right]_m ~\dot{s}_{-1,i\nu}(k_{t}\tau)~
j^m(\tau_2,\nu,{\vec k})~~,
\label{eq:A.20}\end{eqnarray}
i.e., the longitudinal part of the magnetic field has only the
azimuthal component (the magnetic field circulates around the current
flowing in the $\eta$-direction), which is natural for the distribution
of charges which experience expansion in $z$-direction. Note, that
the magnetic field exists even when $\rho$ is $\tau$-independent.

In the same way, we compute the electric fields
\begin{eqnarray}
E^{[L]}_{l}(\tau,\nu,{\vec k})= {\tau \over (2\pi)^3 k_{t}^2}
\left[ \begin{array}{c} k_r  \\ \nu  \end{array} \right]_l
\left[ \begin{array}{c} k_s  Q_{-1,i\nu}(k_{t}\tau) \\
\nu Q_{1,i\nu}(k_{t}\tau_2)
\end{array} \right]_m  j^m(\tau,\nu,{\vec k})~~,
\label{eq:A.21}\end{eqnarray}
and
\begin{eqnarray}
E^{[inst]}_{l}(\tau,\nu,{\vec k})=
{-i\ \over  (2\pi)^3 k_{t}^2}
\bigg\{ \left[ \begin{array}{c} k_r  Q_{-1,i\nu}(k_{t}\tau) \\
\nu Q_{1,i\nu}(k_{t}\tau) \end{array} \right]_l
~\dot{\rho}(\tau,\nu,{\vec k}) \nonumber\\
+ \left[ \begin{array}{c}
k_r \tau^{-1} s_{1,i\nu}(k_{t}\tau) \\
k_t^2 \nu \tau s_{-1,i\nu}(k_{t}\tau)
\end{array} \right]_l  ~\rho(\tau,\nu,{\vec k})\bigg\} ~~.
\label{eq:A.22}\end{eqnarray}
Once again, in the static limit, $j^m=0$, and $\dot{\rho}=0$;
thus, $E^{[L]}=0$ and only the second term in $E^{[inst]}$
survives and becomes the previously found $E^{[stat]}$. Notice
that the time integration in expressions for potentials looks as
retarded, $\tau_1>\tau_2$. This has nothing to do with causal (and
the only one meaningful) retardation. This inequality is due to
the boundary conditions imposed on $A_l$ (to fix the gauge ) when
$A_l$ is being rebuilt from $E_l$, which is simultaneous with the
sources. The same inequality appears when we shall rebuild the charge
dencity $\rho(\tau)$ via $j^m(\tau)$at the previous time.

Now we can move to the fields {\em produced by the currents}, and
leaving the  vanishing effect of $\rho(\tau=0)$ aside. We want
to present the propagator in its general tensor form which
implies that $$A^{[long]}_{l}(x_1)= \int d^4x_2D^{[long]}_{lm}(x_1,x_2)
 j^m(x_2)~~.$$

Let us begin with the  electric fields produced by the component
$j^\eta$ of the current:
\begin{eqnarray}
E^{[L]}_{\eta}(\tau_1,\nu,{\vec k}|j^\eta)= {1 \over (2\pi)^3 }
\bigg[{\tau_1^2\over 2}- \int_{0}^{\tau_1}s_{1,i\nu}(k_{t}t)tdt
\bigg] ~\tau_1 ~j^\eta(\tau_1,\nu,{\vec k}) ~,
\label{eq:A.23}\end{eqnarray}
\begin{eqnarray}
E^{[inst]}_{\eta}(\tau_1,\nu,{\vec k}|j^\eta)={1\over (2\pi)^3 }
\bigg\{-\tau_1~[ 1-s_{1,i\nu}(k_{t}\tau_1)] \int_{0}^{\tau_1}
\tau_2 d\tau_2 j^\eta(\tau_1,\nu,{\vec k})\nonumber\\
-\bigg[{\tau_1^2\over 2}- \int_{0}^{\tau_1}s_{1,i\nu}(k_{t}t)tdt
\bigg] ~\tau_1 ~j^\eta(\tau_1,\nu,{\vec k}) \bigg\} ~.
\label{eq:A.24}\end{eqnarray}
We see, that the $E^{[L]}_{\eta}$ cancel out the second term in
$E^{[inst]}_{\eta}$, originating, in its turn, from the term with
$\dot{\rho}$ in Eq.~(\ref{eq:A.22}). In this way, we obtain the full
form of the
$\eta$-component of the longitudinal field,
\begin{eqnarray}
E^{[long]}_{\eta}(\tau,\nu,{\vec k}|j^\eta)=
\bigg[-\tau +\tau s_{1,i\nu}(k_{t}\tau) \bigg]
\int_{0}^{\tau} {\tau_2 d\tau_2\over (2\pi)^3 }
j^\eta(\tau_2,\nu,{\vec k})~,\nonumber\\
A^{[long]}_{\eta}(\tau_1,\nu,{\vec k}|j^\eta)= \int_{0}^{\tau_1}
{\tau_2 d\tau_2 \over (2\pi)^3 } \bigg[{\tau_2^2-\tau_1^2 \over
2}- \int_{\tau_2}^{\tau_1}s_{1,i\nu}(k_{t}t)t dt \bigg]
j^\eta(\tau_2,\nu,{\vec k})~,
\label{eq:A.25}\end{eqnarray}
where the first term is independent of $\nu$ and ${\vec k}$, and
yields the contact part of the propagator, which (in the
coordinate representation) reads as
\begin{eqnarray}
D^{[contact]}_{\eta\eta}(\tau_1,\tau_2;\eta_1-\eta_2;
\vec{r_1}-\vec{r_2}) =-{\tau_1^2-\tau_2^2 \over 2}\delta
(\eta_1-\eta_2) \delta (\vec{r_1}-\vec{r_2})~.
\label{eq:A.26}\end{eqnarray}

The first line of Eq.~(\ref{eq:A.24}) clearly illustrates its
origin: We started in Eq.~(\ref{eq:A.22}) with the product
$~\nu s_{-1,i\nu}(k_{t}\tau) \rho(\tau,\nu,{\vec k})$. Then, since
$\rho(\tau)$ is developed dynamically, we expressed $\rho$ via
$\partial_\eta j^\eta \to \nu j^\eta$, gaining an extra power of
$\nu$. This allows us to use the relation between  two Lommel
functions, Eq.~(\ref{eq:A.8}), and replace (in fact, after
integrating by parts) $\nu^2 s_{-1,i\nu}\to 1-s_{-1,i\nu}$, which
is equivalent to a straightforward account for the Gauss law. The
$\nu$- and $k_t$-independent unit gives
Eq.~(\ref{eq:A.26}).\footnote{One may wonder, why the same type
contact term does not show up in other dynamics (and gauges, like
$A^0=0$). The propagators of these gauges are constructed in such a
way that the translation invariance and the possibility of a
simple momentum representation are preserved. The price for this
apparent simplicity is the spurious poles in the propagator
without a physically motivated prescription to handle these poles.
These poles reflect an intrinsic uncertainty in the way one can
approach the limit of the static field. In order to fix the gauge
$A^0=0$ completely, one has to impose some boundary condition on
the gauge fields at some time $t$, thus corrupting the translation
invariance and gaining additional terms in the propagator,
which,in fact, are of the same origin as the contact term in the
gauge $A^\tau=0$. At large $\tau_1$ and $\tau_2$, and locally in
the coordinate rapidity $\eta$ (when the curvature of the hypersurface
of the constant
$\tau$ becomes negligible), the gauge $A^\tau=0$  can be
approximated locally by the gauge $A^0=0$ \cite{geg}, provided the
boundary conditions at $\tau=0$ are released. In this domain, the
contribution of the contact term is suppressed by the two small
curvature factors,
$g^{\eta\eta}(\tau_1)g^{\eta\eta}(\tau_2)=\tau_1^{-2}\tau_2^{-2}$.
Therefore, if a usual scattering process between the asymptotic
states takes place at large $\tau$, wedge dynamics will treat
it according to the standard scattering theory.}

The second integral term in Eq.~(\ref{eq:A.25}) can also be Fourier-
transformed into the coordinate representation. We want to do that
here, in order to verify that the contact term is not singled out
artificially, and that it is not canceled  by something hidden in the
second term. To compute the integrals from the function $s_{1,i\nu}$ it
can be conveniently decomposed in the following way,
\begin{eqnarray}
s_{1,i\nu}(x)= S_{1,i\nu}(x)-h_{i\nu}(x)~~,
~~~~~~~~~~~~~~~~\nonumber\\
h_{i\nu}(x)={e^{-\pi\nu /2}\over 2}{\pi\nu /2 \over\sinh(\pi\nu /2
)} [H^{(1)}_{i\nu}(x)+ H^{(2)}_{-i\nu}(x)]~.
\label{eq:A.27}\end{eqnarray}
The function $h_{i\nu}(x)$ obeys the homogeneous Bessel equation,
and thus can describe the field only outside the domain of the source
influence.
In the course of calculations, we use the following integral
representation for the Hankel functions,
\begin{eqnarray}
e^{-\pi\nu/2} H^{{2 \choose 1}}_{\mp i\nu} (k_{t}\tau)= {\pm
i\over \pi} \int_{-\infty}^{\infty} e^{\mp ik_{t}\tau \cosh
\theta} e^{\pm i\nu\theta} d \theta~~.
\label{eq:A.28}
\end{eqnarray}
The Lommel function $S_{1,i\nu}$ has a similar representation,
\begin{eqnarray}
S_{1,i\nu}(x)= x\int_{0}^{\infty}\cosh u \cos\nu u ~e^{-x\sinh
u}du~, \label{eq:A.29}\end{eqnarray}
which allows one to compute the integral $d\nu$ exactly,
\begin{eqnarray}
\int_{-\infty}^{\infty}S_{1,i\nu}(k_{t}\tau)e^{i\nu\eta}d\nu= \pi
k_{t}\tau\cosh\eta e^{-k_{t}\tau\sinh |\eta|} ~~,
\label{eq:A.31}\end{eqnarray}
and from Eq.(\ref{eq:A.28}) it follows,
\begin{eqnarray}
\int_{-\infty}^{\infty} d\nu e^{i\nu\eta}h_{i\nu}(k_{t}\tau) =
\int_{-\infty}^{\infty} d~\theta {\sin [k_{t}\tau\cosh\theta)]
\over \cosh^2(\theta+\eta)} = \int_{-\infty}^{\infty}
d~\theta {\sin [k_{t}\tau\cosh(\theta-\eta)]
\over \cosh^2\theta}  ~~.
\label{eq:A.32}\end{eqnarray}
Next we may write the full Fourier transforms. From
Eq.~(\ref{eq:A.31}), we have
\begin{eqnarray}
\int {d^2 {\vec k}\over (2\pi)^3} e^{i{\vec k}{\vec r}}
\int_{-\infty}^{\infty}S_{1,i\nu}(k_{t}\tau)e^{i\nu\eta}d\nu=
-{\tau\cosh\eta\over 4\pi} \nabla_\bot^2 \int_{0}^{\infty}
~ J_0(kr) e^{-k_{t}\tau\sinh |\eta|}dk \nonumber\\
= -{\tau\cosh\eta\over 4\pi} \nabla_\bot^2
~\bigg[ {1\over({\vec r}^2+\tau^2\sinh^2\eta)^{1/2} }\bigg] ~~.
\label{eq:A.33}\end{eqnarray}
Starting from Eq.~(\ref{eq:A.32}),
we continue by introducing $k_z=k_{t}\sinh\theta$ and
$k_0=k_{t}\cosh\theta=|{\bf k}|$ and changing $d^2{\vec k}d\theta$
for the three-dimensional integration $d^3{\bf k}$. With
$t=\tau\cosh\eta$, ${\bf r}=(x,y, \tau\sinh\eta)$, this leads to
\begin{eqnarray}
\int {d^2 {\vec k}\over (2\pi)^3} e^{i{\vec k}{\vec r}}
\int_{-\infty}^{\infty} d\nu e^{i\nu\eta}h_{i\nu}(k_{t}\tau)
= -\nabla_\bot^2 ~\int {d^2 {\vec k}\over 2i (2\pi)^3}
e^{i{\vec k}{\vec r}} \int_{-\infty}^{\infty} {dk_z\over k_0^3}
[e^{i k_0t-ik_z z}-e^{-i k_0t+ik_z z}]=\nonumber\\
=\int {d^3{\bf k}\over (2\pi)^3}{e^{i{\bf k}{\bf r}}
\over |{\bf k}|^3} \sin k_0 t
= - {\nabla_\bot^2  \over 4\pi }
\bigg[ \theta( r^2-\tau^2){\tau\cosh\eta
\over (r^2+\tau^2\sinh^2\eta)^{1/2}}+ \theta(\tau^2- r^2)\bigg]~.
\label{eq:A.34}\end{eqnarray}
Adding (\ref{eq:A.33}) and (\ref{eq:A.34}) we indeed find that the
$\eta\eta$ component of
the longitudinal propagator vanishes at the distances $r_t$
exceeding $\tau$, i.e. outside the light cone of the position of
the current which creates the field. Finally,
\begin{eqnarray}
E^{[long]}_{\eta}(\tau,\eta_1,{\vec r_1}|j^\eta)=
\int d\eta_2~d{\vec r_2}\int_{0}^{\tau} {\tau_2 d\tau_2\over (2\pi)^3 }
j^\eta(\tau_2,\eta_2,{\vec r_2})
 \bigg\{-\tau \delta (\eta)\delta ({\vec r})~~~\nonumber\\
- {\nabla_\bot^2  \over 4\pi } \bigg[
 \theta( \tau-r_t) \bigg({\tau^2\sinh\eta \over
(r_t^2+\tau^2\sinh^2\eta)^{1/2}}-1\bigg) ~\bigg]~\bigg\}~,
\label{eq:A.35}\end{eqnarray}
where, $\eta=\eta_1-\eta_2$ and ${\vec r}={\vec r_1}-{\vec r_2}$.

The first (contact) term in this formula is indeed very special. It is
not limited by the light-cone boundary. The second term, does have this
boundaries, which are just imposed on the Coulomb-type fields rewritten
in terms of the natural coordinates of wedge dynamics. It also includes
the radiation fields propagating along the light cone $\tau=r_t$.
Therefore, only this term can interfere with the radiation fields of
the transverse modes. This is clear evidence that the cancellation
between the contact term and the non-local parts of the propagator is
impossible. As it was demonstrated in Ref.~\cite{geg}, the transverse
electric field is governed by a usual relativistic wave equation.
Integrating Eq.~(\ref{eq:A.35}) over $\tau$ from  $0$ to $\tau_1$, we
recover the potential, and the $\eta\eta$ component of the propagator,
\begin{eqnarray}
D^{[long]}_{\eta\eta}(\tau_1,\tau_2;\eta,{\vec r})
=-{\tau_1^2-\tau_2^2 \over 2}\delta(\eta) \delta (\vec{r})-
{\nabla_\bot^2\over 4\pi}\int_{\tau_2}^{\tau_1} t dt
\theta(t-r_t)
\bigg[{t\cosh\eta\over R(t)}-1\bigg] ~.
\label{eq:A.36}\end{eqnarray}
The remaining components of the propagator $\Delta^{[long]}_{lm}$
are
\begin{eqnarray}
D^{[long]}_{rs}(\tau_1,\tau_2;\eta,{\vec r})=
{\partial_r\partial_s\over 4\pi}\int_{\tau_2}^{\tau_1} {dt\over t}
\theta(t-r_t)
\bigg[{t\cosh\eta\over R(t)}+1\bigg] ~,
\label{eq:A.37}\end{eqnarray}
\begin{eqnarray}
D^{[long]}_{r\eta}(\tau_1,\tau_2;\eta,{\vec r})=
{\partial_r \over 4\pi} \bigg\{ \theta(t-r_t)
\bigg[\int_{0}^{\tau_1}
{r_t^2 ~\sinh\eta~dt\over R^{3}(t)}-
\int_{\tau_2}^{\tau_1}
{t^2~\sinh\eta~dt\over R^{3}(t)}\bigg] \nonumber\\
+\tanh\eta~\int_{0}^{\tau_2}\delta(t-r_t)dt\bigg\}
= D^{[long]}_{\eta r}(\tau_2,\tau_1;-\eta,{\vec r})~,
\label{eq:A.38}\end{eqnarray}
where $R(t)=[r_t^2+t^2\sinh^2\eta]^{1/2}$.
The propagator identically vanishes at $r_t>\tau$, and the derivatives
of the step-function are confined to the light cone corresponding to
the transient radiation which accompaties the creation of the color
charges.

\section{Appendix B. Subleading terms in forward scattering
of the boost states.}
\label{sec:SNB}
\renewcommand{\theequation}{B.\arabic{equation}}
\setcounter{equation}{0}

In the limit of the forward scattering, the general
formula (\ref{eq:E2.13}), can be rewritten in the following form,
\begin{eqnarray}
M_{1,2\to 1',2'}= {g^2 \over 2^7 \pi}~ \delta(
\nu_1+\nu_2-\nu'_1-\nu'_2) \delta( {\vec k}_1+{\vec k}_2-{\vec
k}'_1-{\vec k}'_2) \nonumber\\
\times \int_{0}^{\infty}\tau_1 d\tau_1
\int_{0}^{\infty}\tau_2 d\tau_2
H^{(1)}_{i\nu'_1} (m'_1\tau_1) H^{(2)}_{i\nu_1}
(m_1\tau_1)H^{(1)}_{i\nu'_2} (m'_2\tau_2) H^{(2)}_{i\nu_2}
(m_2\tau_2) \nonumber\\
\times \bigg[{\nu_1+\nu'_1\over \tau_1^2}~
D^{[00]}_{\eta \eta}(\tau_1,\tau_2;\zeta, {\vec q})
~{\nu_2+\nu'_2\over \tau_2^2}
-~q^r ~D^{[00]}_{rs}(\tau_1,\tau_2;\zeta,{\vec q})
~q^s \nonumber\\
+ {\nu_1+\nu'_1\over \tau_1^2}~
D^{[00]}_{\eta s}(\tau_1,\tau_2;\zeta, {\vec q})~ q^s
- q^r~ D^{[00]}_{r\eta}(\tau_1,\tau_2;\zeta, {\vec q})
~{\nu_2+\nu'_2\over \tau_2^2}\bigg]~,
\label{eq:B.1}\end{eqnarray}
where we took the initial transverse momenta
${\vec p_1}={\vec p_2} =0$, and correspondingly, the final
 state momenta, ${\vec p}'_2=-{\vec p}'_1 = {\vec q}$.
 By its design, the full $T$-ordered propagator $D^{[00]}_{lm}$
 is a sum of the longitudinal part and two terms originating from
 the transverse electric, $V^{(TE)}$, and transverse magnetic,
 $V^{(TM)}$,  modes of the radiation field,
$$D^{[00]}= D^{[00,long]} + D^{[00](TE)} + D^{[00](TM)}.$$

The $\eta\eta$ component of the longitudinal part of the propagator
can be read out from the Eq.~(\ref{eq:A.25}),
\begin{eqnarray}
D^{[long]}_{\eta \eta}(\tau_1,\tau_2;\zeta, {\vec q})
={1\over 2\pi}~\bigg[-{\tau_1^2-\tau_2^2 \over 2}-
\int_{\tau_2}^{\tau_1}s_{1,i\zeta}(q_{t}t)t dt ~\bigg]~,
\label{eq:B.2}\end{eqnarray}
where the first term on the right has already been used
in Eq.~(\ref{eq:E2.15}) to obtain the main estimate (\ref{eq:E2.13e}).
The
second term yields
\begin{eqnarray}
I_1=\int_{\tau_0}^{T}{d\tau_1\over\tau_1}
\int_{\tau_0}^{T}{d\tau_2\over \tau_2}
\bigg({\tau_1\over\tau_2}\bigg)^{-i\zeta}
~{\rm sign}(\tau_1-\tau_2)~ \int_{\tau_2}^{\tau_1}s_{1,i\zeta}(q_{t}t)t
dt ~,
\label{eq:B.3}\end{eqnarray}
The behavior of the Lommel function in the limit $\zeta\to 0$
can be found from the integral representation (\ref{eq:A.19})
\begin{eqnarray}
s_{1,i0}(q_{t}\tau)=
{1\over\pi}\int_0^\pi[1-\cos(k_{t}\tau\sin\phi)]~d\phi
=1-J_0(q_{t}\tau)~~,
\label{eq:B.4}\end{eqnarray}
Expanding the Bessel function $J_0(q_{t}t)$ at small $q_t$,
and integrating, we arrive at
\begin{eqnarray}
2I_1={q_t^2T^4\over \zeta^2+16}
\bigg\{\bigg(1+{\tau_0^4\over T^4}\bigg)~
{\sin[\zeta\ln(T/\tau_0)]\over \zeta}-
\bigg(1-{\tau_0^4\over T^4}\bigg)
{1+\cos[\zeta\ln(T/\tau_0)]\over 4}~\bigg\}~.
\label{eq:B.5}\end{eqnarray}
This term vanishes in the limit of the forward scattering,
$q_t\to 0$.

The $rs$ component of the longitudinal field propagator,
\begin{eqnarray}
D^{[long]}_{rs}(\tau_1,\tau_2;\zeta, {\vec q})
={1\over 2\pi}~ {q_r q_s \over q_t^2}
~\int_{\tau_2}^{\tau_1}s_{1,i\zeta}(q_{t}t)
{dt\over t} ~,
\label{eq:B.6}\end{eqnarray}
brings in the term
\begin{eqnarray}
I_2=q_t^2\int_{\tau_0}^{T}{\tau_1 d\tau_1}
\int_{\tau_0}^{T}{\tau_2 d\tau_2}
\bigg({\tau_1\over\tau_2}\bigg)^{-i\zeta}
~{\rm sign}(\tau_1-\tau_2)~ \int_{\tau_2}^{\tau_1}s_{1,i\zeta}(q_{t}t)
{dt\over t}~.
\label{eq:B.7}\end{eqnarray}
At $\zeta\to 0$ and at small $q_t$, it can be represented
as the integral,
\begin{eqnarray}
2I_2= {q_t^4 T^6 \over 8}~ \int_{\tau_0/T}^{1} y^5
dy \int_{\tau_0/Ty}^{1}
[x^{i\zeta~}+ x^{-i\zeta~}](1-x^2) dx ~,
\label{eq:B.8}\end{eqnarray}
which also vanishes in the limit of the forward scattering,
$q_t\to 0$.

The contribution of the components $D^{[long]}_{r\eta}$ and
$D^{[long]}_{\eta s}$ into the matrix element (\ref{eq:B.1}),
as well as of all components $D^{(TM)}_{lm}$ of the transverse
 propagator is estimated exactly in the same way. All these
  components are defined as the integrals, from $0$ to $\tau q_t$,
of the functions which are regular at the origin. Therefore,
at small $q_t$, all these terms have at least one factor $q_t^2$
and vanish in the limit of the forward scattering.

The only exception from this scheme is the piece connected
with the transverse part $D^{(TE)}$ of the propagator.
This part is the bilinear form $D^{(TE)}_{rs}$ which has only
$rs$-components and includes the projector
$\delta_{rs}- q_r q_s / q_t^2$. Since this projector is
 orthogonal to the vector ${\vec q}_t$, the contribution of this
 mode to the matrix element (\ref{eq:B.1})
identically vanishes.


\begin{references}

\bibitem{gqm}  A. Makhlin, {\em Phys. Rev. C} {\bf 63}, 044902 (2001)
\bibitem{geg}  A. Makhlin, {\em Phys. Rev. C} {\bf 63}, 044903 (2001)
\bibitem{McLerran1} L. McLerran and R. Venugopalan,
{\em Phys. Rev. D} {\bf 49} (1994) 2233; \\{\em Phys. Rev. D} {\bf
49}(1994) 3352;
{\em Phys. Rev. D} {\bf 50} (1994) 2225;\\
A. Ayala, J. Jalilian-Marian, L. McLerran and R. Venugopalan,
{\em Phys. Rev. D} {\bf 52} (1995) 2935.
\bibitem{Kovchegov} Yu. Kovchegov, {\em Phys. Rev. D} {\bf 54} (1996)
5463;\\
Yu. Kovchegov and A. Mueller, {\em Nucl. Phys.} {\bf B529} (1968) 451.
\bibitem{QGD} A. Makhlin, {\em Phys. Rev. C} {\bf 52} (1995) 995.
\bibitem{tev}  A. Makhlin and E. Surdutovich ,
               {\em Phys. Rev. C}{\bf 58} (1998) 389.
\bibitem{McLerran2} J. Jalilian-Marian, A. Kovner, L. McLerran and H.
Weigert,
{\em Phys. Rev. D} {\bf 55} (1997) 5414.
\bibitem{Lam1}S.G. Lam and G. Mahlon, {\em Phys. Rev. D} {\bf 64}
(2001) 016004.
\bibitem{KMW} A. Kovner, J. Milhano and H. Weigert, {\em Phys.Rev.
D}{\bf 62}
(2000) 114005.
\bibitem{Weinberg} S. Weinberg, The quantum theory of fields,
 Cambridge Univ. Press, 1995;\\
R.J. Eden, High energy collisions of elementary particles,
Cambridge Univ. Press, 1967.
\bibitem{Kovner} J. Jalilian-Marian, A. Kovner, A. Leonidov and H.
Weigert,
{\em Phys. Rev. D} {\bf 59} (1999) 014014;
{\em Phys. Rev. D} {\bf 59} (1999) 034007;\\
H. Weigert,  hep-ph/0004044;\\
Yu. Kovchegov, {\em Phys.Rev. D}{\bf 61} (2000) 074018;\\
C.S. Lam, Nucl.Phys.Proc.Suppl. {\bf 79} (1999) 213.
\bibitem{zahed}E.Shuryak, I.Zahed,{\em Phys. Rev. D}{\bf 62} (2000)
085014;\\
M. Nowak, E. Shuryak, I. Zahed ,{\em Phys. Rev. D}{\bf 64} (2001)
034008.
\bibitem{QFK} A. Makhlin, {\em Phys. Rev. C} {\bf 51} (1995) 3454.
\bibitem{fse}  A. Makhlin and E. Surdutovich , {\em Phys. Rev. C}
{\bf 63}, 044904 (2001)
\bibitem{Keld}  L.V. Keldysh, Sov. Phys. JETP {\bf 20} (1964) 1018;
               E.M. Lifshits, L.P. Pitaevsky, Physical kinetics,
             Pergamon Press, Oxford, 1981.
\bibitem{Bjorken} J.D. Bjorken and S.D. Drell, Relativistic quantum
                  mechanicss, McGraw-Hill, New York, 1965.

\end{references}
\end{document}